\documentclass[twocolumn,prb,showpacs,superscriptaddress]{revtex4}
\usepackage{bm}
\usepackage{times}

\begin{document}
\topmargin-1.0cm

\title {Non-adiabatic electron dynamics in time-dependent density-functional theory}

\author {C. A. Ullrich}
\affiliation
{Department of Physics and Astronomy, University of Missouri, Columbia, Missouri 65211}

\author {I. V. Tokatly}
\affiliation
{Lehrstuhl f\"ur Theoretische Festk\"orperphysik, Universit\"at Erlangen-N\"urnberg, Staudtstrasse 7/B2, 91058 Erlangen, Germany}
\affiliation{Moscow Institute of Electronic Technology, Zelenograd, 124498 Russia}
\date{\today}

\begin{abstract}
Time-dependent density-functional theory (TDDFT) treats
dynamical exchange and correlation (xc) via a single-particle potential, $V_{\rm xc}({\bf r},t)$, defined as
a nonlocal functional of the density $n({\bf r}',t')$. The popular adiabatic local-density
approximation (ALDA) for $V_{\rm xc}({\bf r},t)$ uses only densities at the same space-time point $({\bf r},t)$.
To go beyond the ALDA, two local approximations have been proposed based on quantum hydrodynamics and elasticity theory:
(a) using the current as basic variable (C-TDDFT)
[G. Vignale, C. A. Ullrich, and S. Conti, Phys. Rev. Lett. {\bf 79}, 4878 (1997)],
(b) working in a co-moving Lagrangian reference frame (L-TDDFT)
[I. V. Tokatly, Phys. Rev. B {\bf 71}, 165105 (2005)]. This paper illustrates, compares, and analyzes
both non-adiabatic theories for simple time-dependent model densities in the linear and nonlinear regime,
for a broad range of time and frequency scales. C- and L-TDDFT are identical in certain limits, but in
general exhibit qualitative and quantitative differences in their respective treatment of elastic and
dissipative electron dynamics. In situations where the electronic density rapidly undergoes large deformations,
it is found that non-adiabatic effects can become significant, causing the ALDA to break down.
\end{abstract}

\maketitle

\newcommand{\bfrt}{{\bf r},t}
\newcommand{\bfrpt}{{\bf r}',t}
\newcommand{\upalda}{^{\scriptscriptstyle \rm ALDA}}
\section{Introduction} \label{sec:intro}
Time-dependent density-functional theory (TDDFT)  has gained considerable
popularity as a practical approach to the dynamics of many-electron systems.\cite{rungegross,tdft,grodope,rvl}
The essential idea of TDDFT is to describe $N$ interacting electrons moving in an external potential $V({\bf r},t)$
in terms of an auxiliary non-interacting system governed by the
time-dependent Kohn-Sham (TDKS) equation,
\begin{equation}\label{TDKS}
\left[\frac{1}{i} \frac{\partial}{\partial t} -\frac{ \nabla^2}{2}
+ V(\bfrt) + V_{\rm H}(\bfrt)+ V_{\rm xc}(\bfrt)\right]\varphi_\alpha(\bfrt)=0
\end{equation}
(we use atomic units $\hbar = e = m = 1$ throughout).
Here and in the following, we consider systems whose ground
state and dynamical response are everywhere nonmagnetic, and we may therefore ignore the spin degree of freedom.
Eq. (\ref{TDKS}) thus describes the time evolution of doubly occupied Kohn-Sham orbitals $\varphi_\alpha(\bfrt)$,
and the time-dependent density
\begin{equation}\label{nrt}
n(\bfrt) = 2\sum_{\alpha=1}^{N/2} |\varphi_\alpha(\bfrt)|^2
\end{equation}
is obtained in principle exactly.
In Eq. (\ref{TDKS}), $V_{\rm H}(\bfrt) = \int d^3r' n(\bfrpt)/|{\bf r} - {\bf r}'|$ is the time-dependent
Hartree potential, and $V_{\rm xc}({\bf r},t)$ is the exchange-correlation (xc) potential.
In practice, suitable approximations for $V_{\rm xc}(\bfrt)$ are required.
We assume in the following that the system evolves from its ground state at $t=t_0$,
although this assumption is not strictly necessary.

The exact $V_{\rm xc}[n]({\bf r},t)$ has a functional dependence on $n(\bfrt)$ that is nonlocal in
space and time, i.e., contains information about the previous history of the system,
including its initial state. \cite{maitraburke} However, almost
 all present applications of TDDFT employ the adiabatic approximation,
ignoring all functional dependence of $V_{\rm xc}$ on prior time-dependent densities $n({\bf r}',t')$, $t'<t$.
The simplest example is the adiabatic local-density approximation (ALDA):
\begin{equation}\label{alda}
V_{\rm xc}^{\rm ALDA} (\bfrt) = \left. \frac{d\epsilon_{\rm xc}(\bar{n})}{d\bar{n}} \right|_{\bar{n} = n(\bfrt)} ,
\end{equation}
where $\epsilon_{\rm xc}(\bar{n})$ is the xc energy density of a homogeneous electron gas of density $\bar{n}$.
The neglect of retardation in ALDA implies frequency-independent and real xc kernels in linear response. \cite{grosskohn}
This approach has been widely used for calculating molecular excitation energies. \cite{casida,FA02}

One can make the general statement that the adiabatic approximation works well for excitations
of the many-body system that have a direct counterpart in the Kohn-Sham system, such as atomic and
molecular single-particle excitations.\cite{VignaleGiuliani} On the other hand, for more complicated processes
such as double or charge-transfer excitations the ALDA often fails dramatically. \cite{Maitra1,Maitra2}

Several attempts  to go beyond the ALDA can be found in the literature.
\cite{grosskohn,vignalekohn,VUC,ullrichvignale,dobsonbunner,TokatlyPankratov,Tokatly1,Tokatly2,kurzweil}
Vignale and Kohn \cite{vignalekohn} showed that a non-adiabatic {\em local} approximation
for exchange and correlation requires the time-dependent {\em current} ${\bf j}({\bf r},t)$
as basic variable, rather than the density $n({\bf r},t)$ (C-TDDFT).
This formalism was later cast  in a physically more
transparent form using the language of hy\-dro\-dy\-namics, \cite{VUC,ullrichvignale}
where non-adiabatic xc effects appear as viscoelastic stresses in the electron liquid.

To date, C-TDDFT has been applied mainly in fre\-quency-dependent linear response.
The first application was to calculate linewidths of
intersubband plasmons in semiconductor quantum wells. \cite{UllrichVignale1,UllrichVignale2}
In the absence of disorder and phonon scattering, the ALDA gives infinitely sharp plasmon lines.
C-TDDFT includes damping due to electronic many-body effects, in good agreement with experimental
linewidths. \cite{UllrichVignale1,UllrichVignale2}
Van Faassen {\em et al.} \cite{faassen1} calculated static axial polarizabilities
in molecular chains, which are greatly overestimated with ALDA, and achieved
an excellent agreement with ab initio quantum chemical results. Other recent
studies used C-TDDFT to calculate atomic and molecular excitation energies. \cite{ullrichburke,faassen2,berger}

Beyond linear response, a wealth of interesting electron dynamics can be explored using
TDKS theory.\cite{Wijewardane1} The C-TDDFT formalism has recently been applied to describe linear and nonlinear
charge-density oscillations in quantum wells in the time domain.\cite{Wijewardane2}
It was shown that the retardation caused by the memory dependence of the xc potential has the striking
consequence of introducing decoherence and energy relaxation. \cite{Wijewardane2,DAgosta}

An alternative non-adiabatic theory has recently been developed by one
of the authors.  \cite{Tokatly1,Tokatly2} The idea is to relate the
local stress in the electron liquid, and thus the xc potential, to the
dynamics of deformations of fluid elements in the quantum many-body
system.  This leads to a formally exact reformulation of TDDFT from
the point of view of an observer in a co-moving Lagrangian reference
frame (L-TDDFT). Casting the theory in terms of Lagrangian coordinates
allows one to get around the well-known problem of ``ultranonlocality'' in TDDFT, and
to derive, in a rigorous fashion, an exact time-dependent, non-adiabatic extension
of the ground-state LDA.

In the L-TDDFT formalism the xc potential appears as a local
functional of the dynamic deformation tensor. At present, two limiting
forms of this local functional are available. A high-frequency, ``elastic''
form of the non-adiabatic xc potential was derived in
Ref.~\onlinecite{Tokatly2}. The elastic approximation correctly accounts for all
complicated nonlinear deformation effects, but completely neglects
possible xc contributions to dissipation. The second available limiting form of
the non-adiabatic xc potential corresponds to the regime of small
deformations. As we will show, in the limit of small deformations
L-TDDFT formally reduces to C-TDDFT. In this limit the xc stress (and thus the xc potential) is
proportional to the linearized strain tensor which can be considered as a local
linear functional of the current. In fact, C-TDDFT of
Ref.~\onlinecite{VUC} can be viewed as
a viscoelastic linear Hooke's law in the nonlinear quantum continuum
mechanics defined by the general formulation of L-TDDFT. In contrast
to the purely elastic approximation constructed in
Ref.~\onlinecite{Tokatly2}, C-TDDFT contains all dissipation
effects, but is formally restricted to infinitesimally small
deformations.  Of course, the formal
asymptotic criteria and the practical regimes of applicability of any
approximation can be quite different. Therefore a more detailed
analysis based on explicit numerical examples is required to assess
the validity of the two currently available non-adiabatic xc functionals, and
to analyze their relation to ALDA.

Thus, the purpose of this paper is to illustrate, compare, and analyze
C-TDDFT and the elastic approximation to L-TDDFT for simple,
quasi-one-dimensional model systems, in order to show the differences
and common grounds of both approaches. We explore the performance of
the two xc potentials for two kinds of ana\-ly\-ti\-cally given
time-dependent model densities, representing charge-density
oscillations in the form of collective sloshing and breathing modes of
varying amplitudes and frequencies. This will allow us to simulate
electron dynamics over a wide range of time and frequency scales, from
the linear to the nonlinear regime. In our analysis, we will focus on
a detailed comparison of the time-dependent xc potentials, as well as
on the instantaneous and time-averaged power absorption associated
with the charge-density oscillations.  This will give us insight into
the inner workings of C- and L-TDDFT in different dynamical regimes,
and their relation to the ALDA. In particular, we will discuss and
clarify the meaning of ``non-adiabatic'', the cross-over from the
linear to the nonlinear domain, and the competition and coexistence of
elastic and dissipative xc effects.

It turns out that, for small-amplitude deformations, C- and L-TDDFT
agree in the limit of short time scales or, equivalently, in the
high-frequency regime. In general, the two theories show some
differences in their treatment of elastic and dissipative effects in
the inhomogeneous electron liquid.  However, the size of these
differences strongly depends on the type of collective mode, and the
associated charge-density deformations. If the deformations are large
and occur on short time (high frequency) scales, both non-adiabatic
theories give a clear indication of a failure of the ALDA.

This paper is organized as follows. In section \ref{sec:theory} we
summarize the essential formal framework of C- and L-TDDFT. Section
\ref{sec:const} shows how to construct simple analytic model densities
in the Lagrangian and associated laboratory reference frame.  Section
\ref{sec:results} presents detailed numerical results and discussion,
with a separate treatment of the linear and nonlinear regime.  We give
our conclusions in Section \ref{sec:conclusion}.

\section{Non-adiabatic theories in TDDFT} \label{sec:theory}

\subsection{C-TDDFT}

\subsubsection{Linear-response regime} \label{sec:C-TDDFT-linear}

Starting point of C-TDDFT is the linear current-den\-si\-ty response ${\bf j}_1({\bf r},\omega)$ to an external,
frequency-dependent vector potential $\bf A_{\rm ext1} ({\bf r},\omega)$,
\begin{eqnarray} \label{2.A.1}
j_{1,i}({\bf r},\omega) &=& \int \! d^3 r' \chi_{{\rm KS},ij}({\bf r},{\bf r}',\omega)
\left[ A_{{\rm ext1},j}({\bf r}',\omega)\right. \nonumber\\
&& \left. {} + A_{{\rm H1},j}({\bf r}',\omega) +
A_{{\rm xc1},j}({\bf r}',\omega) \right]\:,
\end{eqnarray}
where $i,j$ denote Cartesian coordinates, and $\chi_{{\rm KS},ij}$ is the non-interacting,
Kohn-Sham current-current response tensor. Here and in the following, we use the Einstein convention for
the summation over repeated indices. The Hartree vector potential is given by
\begin{equation} \label{2.A.5}
A_{{\rm H1},j}({\bf r},\omega) = \frac{\nabla_j}{(i\omega)^2}
\int \! d^3 r'  \: \frac{\nabla ' \cdot {\bf j}_1({\bf r}',\omega)}{|{\bf r} - {\bf r}'|} \:.
\end{equation}
The simplest approximation for the linearized xc vector potential ${\bf A}_{\rm xc1}$ is the ALDA, which is defined as
\begin{equation}\label{2.A.6}
A\upalda_{{\rm xc1},j}({\bf r},\omega) =
\frac{\nabla_j}{(i\omega)^2}
\int \! d^3 r'  \: f_{\rm xc}\upalda({\bf r},{\bf r}') \:
\nabla ' \cdot {\bf j}_1({\bf r}',\omega),
\end{equation}
where
\begin{equation} \label{2.A.7}
f_{\rm xc}\upalda({\bf r},{\bf r}') =
\left.\frac{d^2 \epsilon_{\rm xc}(\bar{n})}{d\bar{n}^2}\right|_{\bar{n} = n_0({\bf r})}
\delta({\bf r} - {\bf r}')
\end{equation}
is the frequency-independent ALDA xc kernel [$n_0({\bf r})$ is the ground-state density].
In contrast with the xc scalar potential, the xc vector potential admits a frequency-dependent local approximation.
The resulting expression can be written as follows:\cite{vignalekohn,VUC,ullrichvignale}
\begin{equation}\label{2.A.9}
A_{{\rm xc1},j}({\bf r},\omega) = A\upalda_{{\rm xc1},j}({\bf r},\omega)
- \frac{c}{i\omega n_0({\bf r})} \nabla_k \sigma_{{\rm xc},jk}({\bf r},\omega) \:.
\end{equation}
Here, $c$ denotes the speed of light, and $\sigma_{{\rm xc},jk}$ is the xc viscoelastic stress tensor:
\begin{eqnarray} \label{2.A.10}
\sigma_{{\rm xc},jk} &=& \eta_{\rm xc} \left( \nabla_j v_{1,k} + \nabla_k
v_{1,j} - \frac{2}{3} \nabla \cdot {\bf v}_1 \: \delta_{jk} \right) \nonumber\\
&+& \zeta_{\rm xc}\nabla \cdot {\bf v}_1 \: \delta_{jk} \;,
\end{eqnarray}
where ${\bf v}_1({\bf r},\omega) = {\bf j}_1({\bf r},\omega) / n_0({\bf r})$ is the velocity field associated
with the current response, and
$\eta_{\rm xc}$ and $\zeta_{\rm xc}$ are complex viscosity coefficients  defined as
\begin{equation}\label{2.A.11}
\eta_{\rm xc}(n,\omega) = - \frac{n^2}{i\omega} \: f_{\rm xc}^T (n,\omega)\:,
\end{equation}
\begin{equation}\label{2.A.12}
\zeta_{\rm xc}(n,\omega) = - \frac{n^2}{i\omega} \: \left(f_{\rm xc}^L(n,\omega) - \frac{4}{3} \:
f_{\rm xc}^T(n,\omega)-\frac{d^2 \epsilon_{\rm xc}}{dn^2} \right).
\end{equation}
$f_{\rm xc}^L(n,\omega)$ and $f_{\rm xc}^T(n,\omega)$ are frequency-dependent xc kernels for the
homogeneous electron gas, which can be found in various parametrizations
in the literature. \cite{grosskohn,nifosi,qianvignale} In Eq. (\ref{2.A.10}),
$\eta_{\rm xc}$ and $\zeta_{\rm xc}$ are both evaluated at the local $n_0({\bf r})$.

\subsubsection{Nonlinear regime} \label{sec:C-TDDFT-nonlinear}

The generalization\cite{VUC,Wijewardane2} of C-TDDFT into the nonlinear regime and the time domain
requires solving the following TDKS equation:
\begin{eqnarray}\label{tdks}
i\,\frac{\partial \varphi_\alpha({\bf r},t)}{\partial t}  &=&
\left[ \frac{1}{2}\left(\frac{\nabla}{i} + \frac{1}{c} \,{\bf A}({\bf r},t) +
\frac{1}{c} \,{\bf A}_{\rm xc}({\bf r},t)\right)^2
 \right. \nonumber\\
&&{} + V({\bf r},t) + V_{\rm H}({\bf r},t)\bigg] \varphi_\alpha({\bf r},t) \;.
\end{eqnarray}
Notice that the  Hartree term can be expressed as a scalar potential, and we are free to admit external
scalar as well as vector potentials.

As explained in Ref. \onlinecite{VUC}, the form of the nonlinear xc vector potential is dictated
by a number of general requirements, such as Newton's third law (xc force density follows from a
symmetric stress tensor), and the proper limit in the linear regime, which was discussed in
the subsection above. A formally exact, general expression for ${\bf A}_{\rm xc}$ resulting from
these requirements will be
presented within the Lagrangian framework in Section \ref{LTDDFT}. However, a straightforward
expression for a nonlinear, nonadiabatic xc vector potential, valid up to
second order in the spatial derivatives, follows almost immediately:
\begin{equation}\label{axc}
\frac{1}{c} \,\frac{\partial A_{{\rm xc},i}({\bf r},t)}{\partial t} = - \nabla_i V_{\rm xc}^{\rm ALDA}({\bf r},t)
+  \frac{\nabla_j \sigma_{{\rm xc},ij}({\bf r},t)}{n({\bf r},t)}\:,
\end{equation}
where the viscoelastic stress tensor $\sigma_{\rm xc}$ now contains the time-dependent velocity field
${\bf v}({\bf r},t)={\bf j}({\bf r},t)/n({\bf r},t)$:
\begin{eqnarray}\label{sigmaxc}
\lefteqn{\hspace*{-0.5cm}
\sigma_{{\rm xc},ij}({\bf r},t) = \int_{-\infty}^t dt' \bigg\{
\eta({\bf r},t,t')\bigg[\nabla_i v_j({\bf r},t') + \nabla_j v_i({\bf r},t') } \nonumber\\
&& - \left.\frac{2}{3} \, \nabla \cdot {\bf v}({\bf r},t') \delta_{ij}\right]
+ \zeta({\bf r},t,t') \nabla \cdot {\bf v}({\bf r},t') \delta_{ij} \bigg\}.
\end{eqnarray}
The viscosity coefficients in Eq. (\ref{sigmaxc}) are the Fourier transforms of (\ref{2.A.11}) and (\ref{2.A.12}):
\begin{equation}\label{visct}
\eta({\bf r},t,t') = \left. \int\frac{d\omega}{2\pi} \: \eta(\bar{n},\omega) e^{-i\omega(t-t')}
\right|_{\bar{n} = n({\bf r},t)}
\end{equation}
and similar for $\zeta$.
The apparent ambiguity in Eq. (\ref{visct}) whether the density should be evaluated at $t$ or $t'$
is resolved by noting that the difference involves higher gradient corrections. We emphasize again
that the simple form of Eq. (\ref{axc}) is justified by the basic assumption that the gradients
of the density and velocity are small; the velocity itself, on the other hand, need not be small.
These points will be elaborated in more detail in Section \ref{LTDDFT}, where we will explain
how the approximate expression (\ref{axc}) is obtained from the Lagrangian framework in the appropriate limit.

In the following, we consider model systems where all spatial dependence is along the $x$
direction only. One can then transform the xc vector potential, Eq. (\ref{axc}),  into a scalar one:
$V_{\rm xc}(x,t) =  V_{\rm xc}^{\rm ALDA}(x,t) + V_{\rm xc}^{\rm M}(x,t)$ (ALDA+M),
with the memory part given by
\begin{equation} \label{vxcm}
V_{\rm xc}^{\rm M}(x,t) =  - \int_{-\infty}^x \frac{dx'}{n(x',t)}\:
\nabla_{x'} \, \sigma_{{\rm xc},xx}(x',t) \;.
\end{equation}
Assuming that the system has been in the ground state (with zero velocity field) for $t <0$,
the $xx$ component of the xc stress tensor becomes
\begin{equation}
\sigma_{{\rm xc},xx}(x',t) = \int_0^t Y(n(x',t),t-t') \nabla_{x'} v_{x'}(x',t') dt' \;,
\end{equation}
where the memory kernel $Y$ is given by
\begin{equation}
Y(n,t-t') = \frac{4}{3} \, \eta(n,t-t') + \zeta(n,t-t') \;.
\end{equation}
With the help of the Kramers-Kronig relations for $f_{\rm xc}^L$ we can express the memory kernel as follows:
\begin{equation} \label{Ykernel}
Y(n,t-t') = \frac{4}{3} \,\mu_{\rm xc}
-\frac{n^2}{\pi} \int \frac{d\omega}{\omega} \, \Im f_{\rm xc}^L(\omega) \cos\omega(t-t') ,
\end{equation}
with the xc shear modulus of the electron liquid, \cite{qianvignale}
\begin{equation}
\mu_{\rm xc} = \frac{3n^2}{4} \left( \Re f_{\rm xc}^L(0) - \frac{d^2\epsilon_{\rm xc}}{dn^2}\right) .
\end{equation}
The short-time behavior of $Y(n,t-t')$ is of particular interest, since it governs the high-frequency
dynamics. The limit $Y(n,0)$ can be expressed analytically using
the Kramers-Kronig relation
\begin{equation}\label{A12}
\int\limits_{-\infty}^\infty \frac{d\omega}{\pi} \: \frac{\Im f_{\rm xc}^L(\omega)}{\omega}
= \Re f_{\rm xc}^L(0) - f_\infty^L \:,
\end{equation}
where the high-frequency limit $f_\infty^L$ is known via the third-moment sum rules. \cite{qianvignale}
The result is
\begin{equation}\label{Aini}
Y(n,0) =  -\frac{20}{3} \: \epsilon_{\rm xc} + \frac{26 n }{5} \: \frac{d\epsilon_{\rm xc}}{dn}
- \frac{d^2 \epsilon_{\rm xc}}{dn^2} \:.
\end{equation}
It is also straightforward to see that $\frac{dY}{dt}(n,0)=0$, i.e., the memory kernel starts with zero slope.

\begin{figure}
\unitlength1cm
\begin{picture}(5.0,6.0)
\put(-6.1,-12.6){\makebox(5.0,6.0){\includegraphics{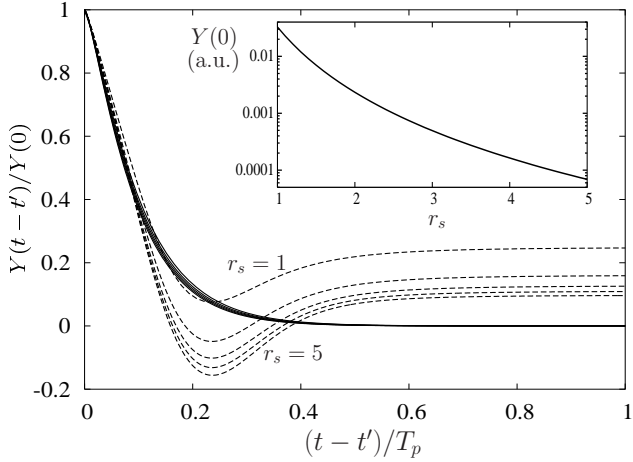}}}
\end{picture}
\caption{\label{figure1} Scaled memory Kernel $Y(n,t-t')$ for $r_s = 1,2,3,4,5$, using the GK (full lines) and
QV (dashed lines) parametrizations.
Inset: short-time limit $Y(n,0)$ [Eq. (\ref{Aini})] versus $r_s$.
}
\end{figure}

Figure \ref{figure1} shows the memory kernel $Y(n,t-t')$ evaluated with the Gross-Kohn (GK) \cite{grosskohn}
and Qian-Vignale (QV) \cite{qianvignale} parametrizations for $f_{\rm xc}^L(\omega)$,
and scaled by the short-time limit $Y(n,0)$ (see inset). Here, $T_p = 2\pi/\omega_p$
is the characteristic time scale associated with the plasma frequency $\omega_p = \sqrt{4\pi n}$. As noted
earlier, \cite{Wijewardane2} $Y^{\rm GK}$ and $Y^{\rm QV}$ are identical at $t=t'$ and have a similar
short-time behavior. $Y^{\rm GK}$ goes exponentially to zero for large $t-t'$, and
the scaled GK memory kernels are nearly identical for all times over a wide range of $r_s$. On the other hand,
all $Y^{\rm GK}$ pass through a minimum around $0.22 T_p$ and then approach the finite limit $4\mu_{\rm xc}/3$,
where $\mu_{\rm xc}\to 0$ for large $r_s$. \cite{qianvignale}
It is interesting to notice that both QV and GK memory kernels
reach their long-time asymptotic limits very rapidly, within about half a plasma cycle $T_p$.

\subsection{L-TDDFT} \label{LTDDFT}
\subsubsection{General formulation}

The main problem in constructing nonadiabatic approximations for xc
potentials is an inherent nonlocality of the time-dependent
theory. Physically, this nonlocality is related to the convective
motion of the electron fluid (the particles at a given point of
space retain the memory of their previous positions).\cite{Tokatly2}
The key idea of L-TDDFT is to eliminate the above
source of nonlocality by reformulating the theory in the Lagrangian
frame, i.e., in a local reference frame moving with the fluid. Since
the convective motion in the Lagrangian frame is absent, a spatially local
description of xc effects becomes possible. This possibility represents
the most important outcome of L-TDDFT: it allows one to derive an exact
nonadiabatic extension of LDA into the dynamic regime.

The general formulation of L-TDDFT starts with the exact relation of
the xc force to the xc stress tensor $P_{{\rm xc},ij}$.
\cite{TokatlyPankratov,Tokatly2} By definition the xc potential
${\bf A}_{\rm xc}$ ensures that the physical density and current are
reproduced by an auxiliary system of noninteracting KS particles. This
means that ${\bf A}_{\rm xc}$ should produce an
effective xc Lorentz force that exactly compensates for a difference of local
stress forces in the real interacting system and in the noninteracting
KS system. Accordingly, the xc vector potential should satisfy the
following equation:
\begin{equation}
-\frac{\partial A_{{\rm xc},i}}{\partial t} + v_{j}(\nabla_{i}A_{{\rm xc},j} -
\nabla_{j}A_{{\rm xc},i}) = \frac{c}{n}\nabla_{j}P_{{\rm xc},ij} \:,
\label{L1}
\end{equation}
where  $P_{{\rm xc},ij} = P_{ij} - T^{\rm KS}_{ij}$ is the difference of
the full stress tensor for the interacting system, $P_{ij}$, and the
kinetic stress tensor for KS system, $T^{\rm KS}_{ij}$. Equation (\ref{L1})
serves as a basic definition of ${\bf A}_{\rm xc}$, which automatically
accounts for the zero force and zero torque conditions.\cite{vignalekohn}

We note that the xc stress tensor $P_{{\rm xc},ij}$ is to be
distinguished from the earlier introduced xc stress tensor
$\sigma_{{\rm xc},ij}$. The connection between the two will be
explained in detail in subsection \ref{LTDDFTsmall}.  The main
difference lies in the fact that $P_{{\rm xc},ij}$, formally exactly
and to all orders in the inhomogeneity, accounts for {\em all}
dynamical xc effects, whereas the ALDA has been separated out in the
definition of $\sigma_{{\rm xc},ij}$.  Furthermore, $\sigma_{{\rm
xc},ij}$ is valid only for small deformations of the electron liquid
(in a sense to be defined below).

All the rest of L-TDDFT can be viewed as a calculation of the dynamic
xc stress tensor $P_{{\rm xc},ij}$, which enters the definition of ${\bf
  A}_{\rm xc}$, Eq.~(\ref{L1}), by
reformulating the problem in the co-moving Lagrangian frame.
The transformation to the Lagrangian frame corresponds to a
nonlinear transformation of coordinates, ${\bf r}={\bf
r}({\bm\xi},t)$, where ${\bf r}({\bm\xi},t)$ is the trajectory
of an infinitesimal fluid element that evolves from the point
$\bm\xi$. Formally the function ${\bf r}({\bm\xi},t)$ is defined by the
following initial value problem:
\begin{equation}
\frac{\partial {\bf r}({\bm\xi},t)}{\partial t} =
{\bf v}({\bf r}({\bm\xi},t),t), \quad {\bf r}({\bm\xi},0) = \bm\xi \:.
\label{L2}
\end{equation}
The initial positions, $\bm\xi$, of the fluid elements play the role of
spatial coordinates in the co-moving frame. The above transformation
from the old coordinates ${\bf r}$ to the new coordinates $\bm\xi$
induces a change of metric. The metric tensor in the
Lagrangian $\bm\xi$-space coincides with the Green's deformation
tensor,
\begin{equation}
g_{ij}({\bm\xi},t) =
\frac{\partial r_{k}({\bm\xi},t)}{\partial\xi_{i}}
\frac{\partial r_{k}({\bm\xi},t)}{\partial\xi_{j}} \:,
\label{L3}
\end{equation}
which is a common characteristic of deformations in the
Lagrangian formulation of continuum mechanics.
A complete reformulation of TDDFT in the co-moving frame shows that a
local description of xc effects is possible if one uses
$g_{ij}({\bm\xi},t)$ as a basic variable: the stress tensor
in the Lagrangian frame can be consistently considered as a spatially local
functional of the Green's deformation tensor.

For most practical applications we need the xc potential in the laboratory
frame, Eq.~(\ref{L1}). Transforming the stress tensor functional from the
co-moving to the laboratory frame, we find the required tensor
$P_{{\rm xc},ij}$
entering Eq.~(\ref{L1}). The locality in the Lagrangian frame
translates to a local dependence of $P_{{\rm xc},ij}$ on the
Cauchy's deformation tensor,
\begin{equation}
\bar{g}_{ij}({\bf r},t) =
\frac{\partial \xi_{k}({\bf r},t)}{\partial r_{i}}
\frac{\partial \xi_{k}({\bf r},t)}{\partial r_{j}} \:,
\label{L4}
\end{equation}
where the function $\bm\xi({\bf r},t)$ is obtained by inverting the
trajectory equation, ${\bf r}={\bf r}({\bm\xi},t)$. The Cauchy's tensor
$\bar{g}_{ij}({\bf r},t)$ is commonly used to describe deformations in
the Eulerian formulation of continuum mechanics. From a physical point of view, the functional
dependence of the stress tensor on $\bar{g}_{ij}$ thus emerges quite naturally.

Substitution of $P_{{\rm xc},ij}[\bar{g}_{ij}]$ into
Eq.~(\ref{L1})  yields the xc vector
potential ${\bf A}_{\rm xc}$ as a functional of the Cauchy's deformation
tensor. If gradients of the deformation tensor are small, the functional
$P_{{\rm xc},ij}[\bar{g}_{ij}]$ is local in space, but in general it
can be nonlocal in time. The time-nonlocality of the stress tensor
functional is related to the memory loss due to dissipation. In the
absence of dissipation (e.~g. in the exchange-only approximation) the
stress tensor becomes a simple function of $\bar{g}_{ij}$, which corresponds
to an infinitely long memory.

Equations (\ref{L2}) and (\ref{L4}) determine the Cauchy's deformation
tensor as a functional of velocity ${\bf v}({\bf r},t)$. An
alternative, and practically more convenient way to compute
$\bar{g}_{ij}$ for a given velocity is to solve the equation of
motion that governs the dynamics of $\bar{g}_{ij}({\bf r},t)$
directly in the laboratory frame:
\begin{equation}
\frac{\partial\bar{g}_{ij}}{\partial t}
+ v_{k}\frac{\partial \bar{g}_{ij}}{\partial r_{k}} =
- \frac{\partial v_{k}}{\partial r_{i}}\bar{g}_{kj}
- \frac{\partial v_{k}}{\partial r_{j}}\bar{g}_{ik}, \quad
 \bar{g}_{ij}({\bf r},0) = \delta_{ij} \:.
\label{L5}
\end{equation}
An important property of the deformation tensor is that it allows us to
relate the time-dependent density $n({\bf r},t)$ to the initial
density distribution, $n_{0}({\bf r})$:
\begin{equation}
n({\bf r},t) = \sqrt{\bar{g}({\bf r},t)}n_{0}(\bm\xi({\bf r},t)),
\label{L6}
\end{equation}
where $\bar{g}({\bf r},t)$ is the determinant of $\bar{g}_{ij}({\bf r},t)$.

Equation (\ref{L5}) or, equivalently, Eqs.~(\ref{L2}) and (\ref{L4}),
show that in general the deformation tensor is a strongly nonlocal
(both in space and in time) functional of the velocity. Therefore,
in spite of the fact that the xc stress tensor and consequently the xc vector potential
are local functionals of $\bar{g}_{ij}$, they are nonlocal in
terms of velocity or any other variable. This makes the
choice of $\bar{g}_{ij}({\bf r},t)$ as a basic variable much more
preferable.

To obtain an explicit construction of the local functional $P_{{\rm
xc},ij}[\bar{g}_{ij}]$, a solution of a homogeneous
time-dependent many-body problem in the Lagrangian frame is required (see Sec.~V
in Ref.~\onlinecite{Tokatly2}). A complete solution of this problem
seems to be impossible, at least at the current level of knowledge.
However, there are two practically important, exactly solvable special
cases, which are described below.

\subsubsection{Small deformation approximation: Recovery of C-TDDFT} \label{LTDDFTsmall}

The abovementioned many-body problem in the homogeneously deformed Lagrangian
$\bm\xi$-space can be solved perturbatively if the deformation tensor
$\bar{g}_{ij}$ only slightly deviates from the unit tensor $\delta_{ij}$:
\begin{equation}
\bar{g}_{ij}({\bf r},t) = \delta_{ij} + \delta\bar{g}_{ij}({\bf r},t) \:.
\label{L7}
\end{equation}
Introducing the displacement vector,
${\bf u}({\bf r},t) = {\bf r} - {\bm\xi}({\bf r},t)$, and using
Eq.~(\ref{L4}), we find that small $\delta\bar{g}_{ij}$ corresponds to
small gradients of the displacement:
\begin{equation}
\delta\bar{g}_{ij}({\bf r},t) =
- \left(\frac{\partial u_{i}}{\partial r_{j}} +
\frac{\partial u_{j}}{\partial r_{i}}\right) \:.
\label{L8}
\end{equation}
Clearly, small gradients of ${\bf u}({\bf r},t)$ imply that the velocity
gradients are also small, since to lowest order in
$\nabla_{i}u_{j}$ Eq.~(\ref{L2}) reduces to the relation
$\partial_t {\bf u}({\bf r},t)={\bf v}({\bf r},t)$. Obviously the smallness of
deformations does not mean that the displacement or the velocity themselves are
small (i.e., the system can be far beyond the linear response regime). A well
known example is the rigid motion of a many-body system in a harmonic
potential, where $\bar{g}_{ij}=\delta_{ij}$ but the displacement can be
arbitrarily large.

The stress tensor functional for small displacement vectors was derived in
Ref.~\onlinecite{Tokatly2}. Extension of this derivation to the
general regime of small deformations, i.e., to the regime of small
displacement gradients, is straightforward. The resulting
xc stress tensor takes the following form:
\begin{equation}
P_{{\rm xc},ij}({\bf r},t)
= P_{\rm xc}^{\rm ALDA}\big(n({\bf r},t)\big)\delta_{ij}
+ \delta P_{ij}({\bf r},t) \:,
\label{L9}
\end{equation}
where $P_{\rm xc}^{\rm ALDA}(n)$ is the xc pressure of a homogeneous
electron gas, and $\delta P_{ij}$ is a nonadiabatic correction,
which is linear in $\delta \bar{g}_{ij}$:
\begin{eqnarray} \nonumber
&&\delta P_{ij}({\bf r},t) = \int\limits_{0}^{t}dt'\bigg\{
\frac{\delta_{ij}}{2}\widetilde{K}_{\rm xc}\big(n({\bf r},t),t-t'\big)
\delta\bar{g}_{kk}({\bf r},t')  \\
&+& \mu_{\rm xc}\big(n({\bf r},t),t-t'\big)
\Big[ \delta\bar{g}_{ij}({\bf r},t')
- \frac{\delta_{ij}}{3}\delta\bar{g}_{kk}({\bf r},t')\Big]
\bigg\}.
\label{L10}
\end{eqnarray}
The kernels $\mu_{\rm xc}(n,t-t')$ and $\widetilde{K}_{\rm
xc}(n,t-t')$ in Eq.~(\ref{L10}) have the meaning of nonadiabatic shear
and bulk moduli, respectively [the adiabatic part of the bulk modulus
is included in the ALDA pressure term in Eq.~(\ref{L9})]. The
corresponding Fourier transforms of the elastic moduli, $\mu_{\rm
xc}(n,\omega)$ and $\widetilde{K}_{\rm xc}(n,\omega)$, are related to
the complex viscosity coefficients of Eqs.~(\ref{2.A.11}) and (\ref{2.A.12}),
$\eta_{\rm xc}(n,\omega)$ and $\zeta_{\rm xc}(n,\omega)$,
as follows:
\begin{equation}
\mu_{\rm xc}(\omega) = -i\omega\eta_{\rm xc}(\omega),
\quad
\widetilde{K}_{\rm xc}(\omega) = -i\omega\zeta_{\rm xc}(\omega) \:.
\label{L11}
\end{equation}
Using Eq.~(\ref{L11}) and the relation $\partial_t{\bf u}={\bf v}$, we find
that the nonadiabatic stress tensor $\delta P_{ij}$, Eq.~(\ref{L10}),
is identical, up to a sign, \cite{footnote1} to the tensor
$\sigma_{{\rm xc},ij}$ of
Eq.~(\ref{sigmaxc}) [i.~e. $\delta P_{ij}=-\sigma_{{\rm xc},ij}$]. In
addition, in the limit of small
displacement/velocity gradients the spatial derivatives of ${\bf
A}_{\rm xc}$ on the left hand side of Eq.~(\ref{L1}) are
negligible. Thus, in the regime of small deformations we recover the
complete nonlinear form of C-TDDFT, \cite{VUC,Wijewardane2} Eqs.~(\ref{axc}) and
(\ref{sigmaxc}).

The imaginary parts of the complex elastic moduli,
$\widetilde{K}_{\rm xc}(\omega)$ and $\mu_{\rm xc}(\omega)$, are
responsible for the dissipation (viscous) effects. For the
high-frequency/short-time dynamics these effects become irrelevant. As
a result, the high-frequency limit of the nonadiabatic stress tensor of Eq.~(\ref{L10}) becomes
completely local and purely elastic:
\begin{eqnarray}\nonumber
\lefteqn{
\delta P_{ij}^\infty({\bf r},t) =
\frac{\delta_{ij}}{2}\widetilde{K}_{\rm xc}^{\infty}\big(n({\bf r},t)\big)
 \delta\bar{g}_{kk}({\bf r},t)}\\
&+& \mu_{\rm xc}^{\infty}\big(n({\bf r},t)\big)
\Big[ \delta\bar{g}_{ij}({\bf r},t)
- \frac{\delta_{ij}}{3}\delta\bar{g}_{kk}({\bf r},t)\Big] \:,
\label{L12}
\end{eqnarray}
where $\widetilde{K}_{\rm xc}^{\infty}(n)$ and $\mu_{\rm
  xc}^{\infty}(n)$ are the high-frequency limits of the bulk and shear
moduli, respectively.

The structure of the small deformation approximation, Eqs.~(\ref{L9})
and (\ref{L10}), clearly demonstrates that in this regime the
nonadiabatic contribution appears as a small, linear in $\delta\bar{g}_{ij}$
correction to the adiabatic dynamics. If the process is strongly
nonadiabatic, the deformations can not be considered small. In fact,
the deviation of the deformation tensor from $\delta_{ij}$
can serve as a general measure of nonadiabaticity.

\subsubsection{Nonlinear elastic approximation to L-TDDFT}

It is very difficult to account both for the full
nonlinear dependence on $\bar{g}_{ij}$, and for the dissipation.
C-TDDFT includes all xc dissipation effects on a level linear in
$\delta\bar{g}_{ij}$. On the contrary, if
we neglect the dissipation effects, a closed nonlinear local
approximation for the stress tensor can be rigorously derived.
\cite{Tokatly2} The reason is that
the homogeneous many-body problem, which has been formulated in
Ref.~\onlinecite{Tokatly2}, admits a simple complete solution in the regime
of fast dynamics when the dissipation is irrelevant. In
this case the xc stress tensor $P_{{\rm xc},ij}({\bf r},t)$ becomes a
function of the time-dependent density $n({\bf r},t)$ and the Cauchy's
deformation tensor $\bar{g}_{ij}({\bf r},t)$:
\begin{equation}
P_{{\rm xc},ij} = \frac{2}{3}\bar{g}_{ij}\sqrt{\bar{g}}
E_{\rm kin}^{\rm xc}\left(\frac{n}{\sqrt{\bar{g}}}\right)
+ L_{ij}(\bar{g}_{kl})
E_{\rm pot}\left(\frac{n}{\sqrt{\bar{g}}}\right),
\label{L13}
\end{equation}
where $E_{\rm kin}^{\rm xc}(n)$ and $E_{\rm pot}(n)$ are the xc
kinetic energy and the potential energy per unit volume of the
homogeneous electron gas.  The function
$L_{ij}(\bar{g}_{kl})$ in Eq.~(\ref{L13}) is explicitly
defined in Appendix~A of Ref.~\onlinecite{Tokatly2}.
In the limit of small deformations the nonlinear elastic approximation
of Eq.~(\ref{L13}) can be expanded around $\bar{g}_{ij}=\delta_{ij}$ and reduces to the linearized form defined by
Eqs.~(\ref{L9}) and (\ref{L12}). In other words one recovers the
high-frequency limit of C-TDDFT.

We conclude this section with the explicit formulation of the nonlinear
elastic approximation for a one-dimensional motion. If all spatial
variations are along the $x$-axis only, the deformation tensor takes a
diagonal form with $\bar{g}_{zz}=\bar{g}_{yy}=1$, and
$\bar{g}_{xx}=\bar{g}(x,t)$. The xc effects can then be described by an xc
scalar potential that is related to the xc stress tensor as follows:
\begin{equation}\label{vxce}
V_{\rm xc}^{\rm E}(x,t) = \int\limits_{-\infty}^{x}\frac{d x'}{n(x',t)}
\frac{\partial}{\partial x'}P_{{\rm xc},xx}\big(n(x',t),\bar{g}(x',t)\big).
\label{L14}
\end{equation}
Equation (\ref{L13}) for the $xx$-component of the xc stress tensor
reduces to the form
\begin{equation}
P_{{\rm xc},xx}(n,\bar{g}) = \frac{2}{3}\bar{g}^{3/2}
E_{\rm kin}^{\rm xc}\left(\frac{n}{\sqrt{\bar{g}}}\right)
+ L(\bar{g})E_{\rm pot}\left(\frac{n}{\sqrt{\bar{g}}}\right),
\label{L15}
\end{equation}
where the factor $L(\bar{g})$ is given by
\begin{equation}
L(\bar{g}) = \frac{\bar{g}}{\bar{g}-1}\left[ 1
- \frac{\arctan\sqrt{\bar{g}-1}}{\sqrt{\bar{g}-1}}\right].
\label{L16}
\end{equation}
Finally, Eq.~(\ref{L5}), which relates the deformation $\bar{g}(x,t)$
to the velocity $v(x,t)$, simplifies as follows:
\begin{equation}
\frac{\partial\bar{g}}{\partial t} =
- v\frac{\partial \bar{g}}{\partial x}
- 2\frac{\partial v}{\partial x}\bar{g}, \quad
\bar{g}(x,0) = 1.
\label{L17}
\end{equation}
It is worth mentioning that $L(\bar{g}\to 1)=1/3$ in the limit of zero deformation. Eq.~(\ref{L15})
then approaches the standard virial expression for the xc pressure, and
$V_{\rm xc}^{\rm E}(x,t)$ reduces to the ALDA xc potential.
We define the post-ALDA contribution of the nonlinear elastic L-TDDFT xc potential as
\begin{equation} \label{vxce-tilde}
\widetilde{V}_{\rm xc}^{\rm E}(x,t) =
V_{\rm xc}^{\rm E}(x,t) - V_{\rm xc}^{\rm ALDA}(x,t) \:.
\end{equation}
To shorten the notation for the rest of this paper: from now on, whenever we refer to L-TDDFT,
we actually mean the nonlinear elastic approximation to L-TDDFT, defined in this section.

\section{Construction of One-dimensional analytical examples} \label{sec:const}

In one dimension, the trajectory of a fluid element with Lagrangian coordinate
$\xi$ is determined by the following equation:
\begin{equation} \label{traj:eq}
\frac{\partial x(\xi,t)}{\partial t} = v\Big(x(\xi,t),t\Big)\:, \qquad x(\xi,0) = \xi \:.
\end{equation}
$x$ is the position, at time $t$, of a fluid element that started out at $\xi$, and $v$ is its velocity.
In general, this is a complicated nonlinear differential equation, with formal solution
\begin{equation}\label{traj1}
x(\xi,t) = \xi + \int\limits_0^t v(\xi,t') dt' \:.
\end{equation}
In other words: if we know the velocity, at time $t$, of a fluid element that started out at $\xi$, we can
determine its trajectory by direct integration. From this, we can then determine the
time-dependent density in the laboratory frame: we first invert  (\ref{traj1}) to obtain
$\xi(x,t)$, then compute the deformation  as
\begin{equation} \label{gdef}
\bar{g}(x,t) = \left(\frac{\partial \xi}{\partial x}\right)^2 ,
\end{equation}
and finally obtain
\begin{equation}\label{denLab}
n(x,t) = \sqrt{\bar{g}(x,t)} \: n_0\Big(\xi(x,t)\Big) \:.
\end{equation}
In practice, of course, this procedure is not very helpful, since the functional
form of $v(\xi,t)$ is unknown. However, we can use it to construct simple analytic examples, as follows.

\subsection{Sloshing mode}

We assume that the system is confined within hard walls, $-L/2 \le (x,\xi) \le L/2$, with initial density
\begin{equation} \label{n0}
n_0(\xi) = \frac{2N}{L} \: \cos^2\left( \frac{\pi \xi}{L}\right)\:,
\end{equation}
where $N$ is the number of electrons per unit area (sheet density) in the $y-z$ plane.
We assume a simple quadratic form of the velocity field:
\begin{equation}\label{vel}
v(\xi,t) = A \omega\left(\frac{L}{4} - \frac{\xi^2}{L}\right) \cos \omega t \:.
\end{equation}
Equation (\ref{traj1}) is then easily integrated:
\begin{equation}\label{traj2}
x(\xi,t) = \xi + A \: \left(\frac{L}{4} - \frac{\xi^2}{L}\right) \sin \omega t \:.
\end{equation}
The next step is to invert Eq. (\ref{traj2}) to determine the trajectories of the fluid elements, which requires
solving a quadratic equation, with the result
\begin{equation}\label{xi}
\xi(x,t) = \frac{L}{2 A \sin\omega t}
\left( 1 - \sqrt{ 1 + A^2 \sin^2\omega t - \frac{4Ax}{L} \: \sin\omega t } \right)
\end{equation}
which properly reduces to $\xi = x$ for $A\to 0$. The range of allowed
amplitudes is $|A| \le 1 $, which is dictated by the constraint that no fluid element can cross the hard-wall boundaries at
$\pm L/2$.

We can now calculate the deformation using Eq. (\ref{gdef}):
\begin{equation} \label{gsl}
\bar{g}(x,t)
= \left( 1 + A^2 \sin^2\omega t - \frac{4Ax}{L} \: \sin\omega t\right)^{-1}.
\end{equation}
The time-dependent density of the sloshing mode in the laboratory frame, $n(x,t)$, then follows from Eq. (\ref{denLab}),
using (\ref{xi}) and (\ref{gsl}).

\begin{figure}
\unitlength1cm
\begin{picture}(5.0,9.5)
\put(-8.,-11.){\makebox(5.0,9.5){\includegraphics{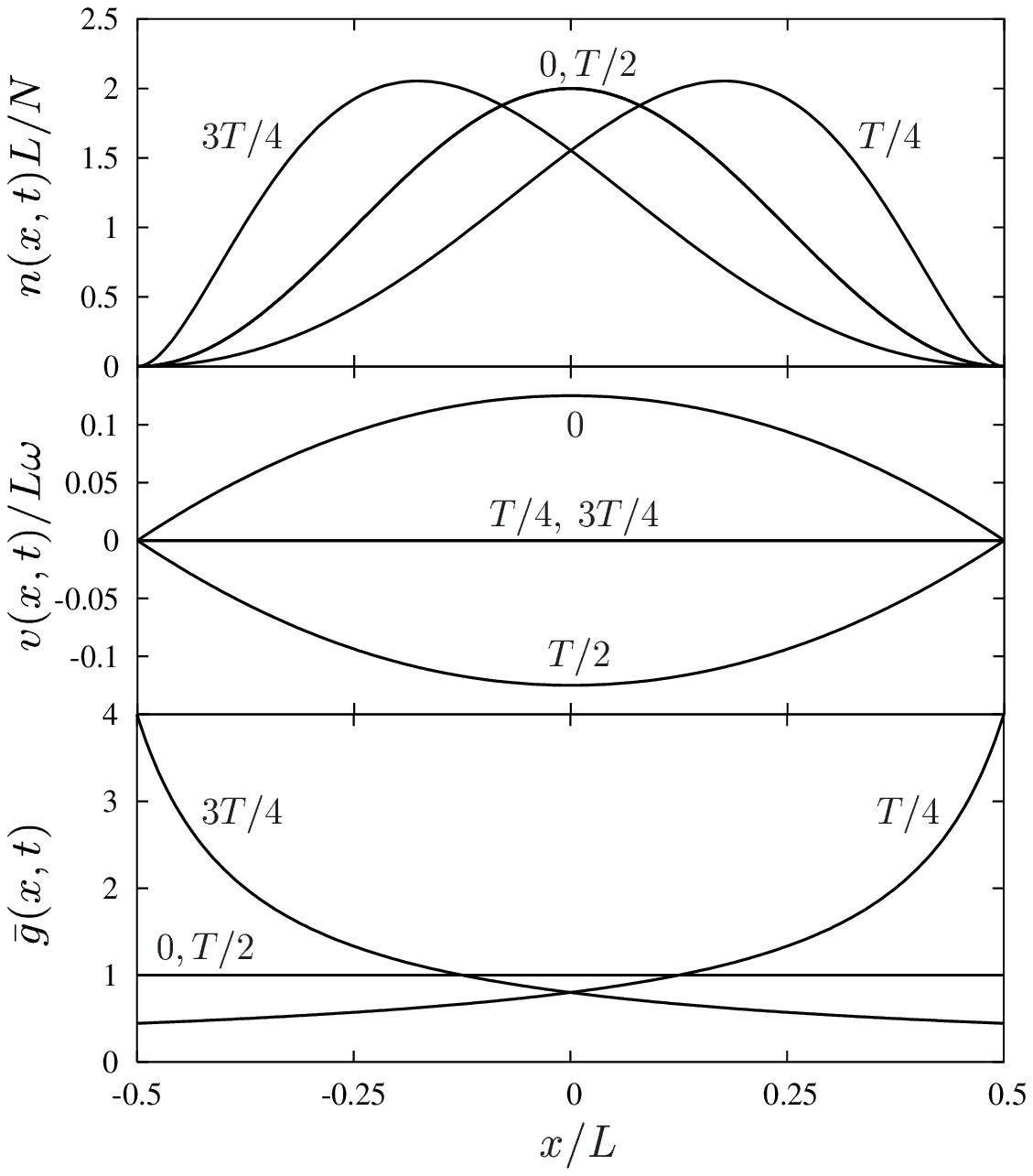}}}
\end{picture}
\caption{\label{figure2} Snapshots of density $n(x,t)$, in units of $N/L$, velocity $v(x,t)$, in units of $L\omega$,
and deformation $\bar{g}(x,t)$ for the sloshing mode
in the laboratory frame, taken at times $t=0, T/4, T/2, 3T/4$.}
\end{figure}

\begin{figure}
\unitlength1cm
\begin{picture}(5.0,9.5)
\put(-8.,-11.){\makebox(5.0,9.5){\includegraphics{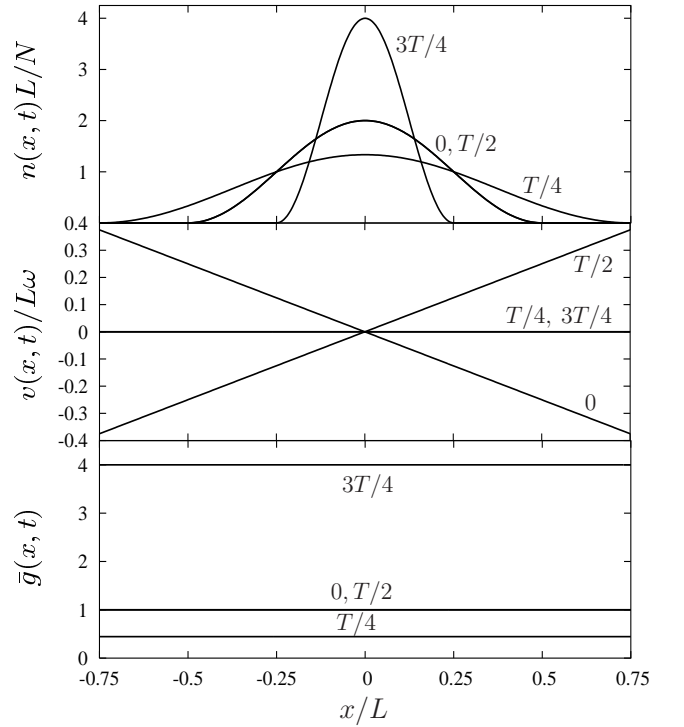}}}
\end{picture}
\caption{\label{figure3} Same as Fig. \ref{figure2}, for the breathing mode.}
\end{figure}

\subsection{Breathing mode}

To simulate a breathing mode, we assume a linear velocity distribution of the fluid elements:
\begin{equation}
v(\xi,t) = A \omega \xi \cos\omega t \:.
\end{equation}
According to Eq. (\ref{traj1}), this gives the following trajectory:
\begin{equation}
x (\xi,t) = \xi(1 + A  \sin\omega t ) \;, \qquad |A|<1 \:.
\end{equation}
This is easily inverted:
\begin{equation}
\xi(x,t) = \frac{x}{1+A\sin\omega t} \:,
\end{equation}
and the resulting deformation is
\begin{equation}
\bar{g}(x,t) = \frac{1}{(1+A\sin\omega t)^2} \:.
\end{equation}
We choose the same initial density distribution $n_0(\xi)$, Eq. (\ref{n0}), as for the sloshing mode,
and the resulting time-dependent density of the breathing mode is
\begin{equation}
n(x,t) = \frac{2N}{L}  \cos^2\left(\frac{\pi x/L}{1 + A\sin\omega t}\right)(1 + A\sin\omega t)^{-1},
\end{equation}
where $|x| < (L/2)(1 + A\sin\omega t)$.

Figs. \ref{figure2} and \ref{figure3} show snapshots of $n(x,t)$, $v(x,t)$, and $\bar{g}(x,t)$ for the sloshing and the
breathing mode, taken at time $t=0, T/4, T/2, 3T/4$, where $T=2\pi/\omega$. The amplitude is $A=0.5$ in both cases,
length is measured in units of $L$, and density and velocity are plotted in units of $N/L$ and $L\omega$, respectively.
The deformation $\bar{g}(x,t)$ is maximal at the turning points of the oscillations (at $t=T/4$ and $3T/4$). We find
that the breathing mode features large deformations everywhere, i.e., $\bar{g}$ deviates strongly from one.
The sloshing mode, on the other hand, is strongly deformed only towards the edges where the density is small.
We will see below how this affects the non-adiabaticity of the xc potential of the two modes.

\section{Results and Discussion} \label{sec:results}

In the following numerical examples, we choose a system where, in atomic units, we have sheet density $N=1$ a.u. and
quantum well width $L=10$ a.u. We shall measure frequencies in units of the average plasmon frequency of the
system, given by
\begin{equation}
\bar{\omega}_p = \frac{1}{L} \int dx\:\omega_p(n(x))\:.
\end{equation}
For our initial density distribution $n_0(x)$, Eq. (\ref{n0}), we find
$\bar{\omega}_p = \sqrt{32 N/\pi L}$, which for the above values of $N$ and $L$ comes out
as $\bar{\omega}_p = 1.009$ a.u.

\subsection{Linear regime}

\subsubsection{Time-dependent xc potentials}

We first consider the linear regime of small density fluctuations.
For the time-dependent densities associated with the breathing mode and the sloshing
mode, we calculate and compare three different time-dependent xc potentials:
$V_{\rm xc}^{\rm M}(x,t)$ from C-TDDFT [Eq. (\ref{vxcm})], $\widetilde{V}_{\rm xc}^{\rm E}(x,t)$ from L-TDDFT
[Eq. (\ref{vxce-tilde})], and the ALDA potential fluctuations $\widetilde{V}_{\rm xc}^{\rm ALDA}(x,t) =
V_{\rm xc}^{\rm ALDA}(n(x,t)) - V_{\rm xc}^{\rm LDA}(n_0(x))$.

\begin{figure}
\unitlength1cm
\begin{picture}(5.0,13.25)
\put(-9.25,-10.25){\makebox(5.0,13.25){
\includegraphics{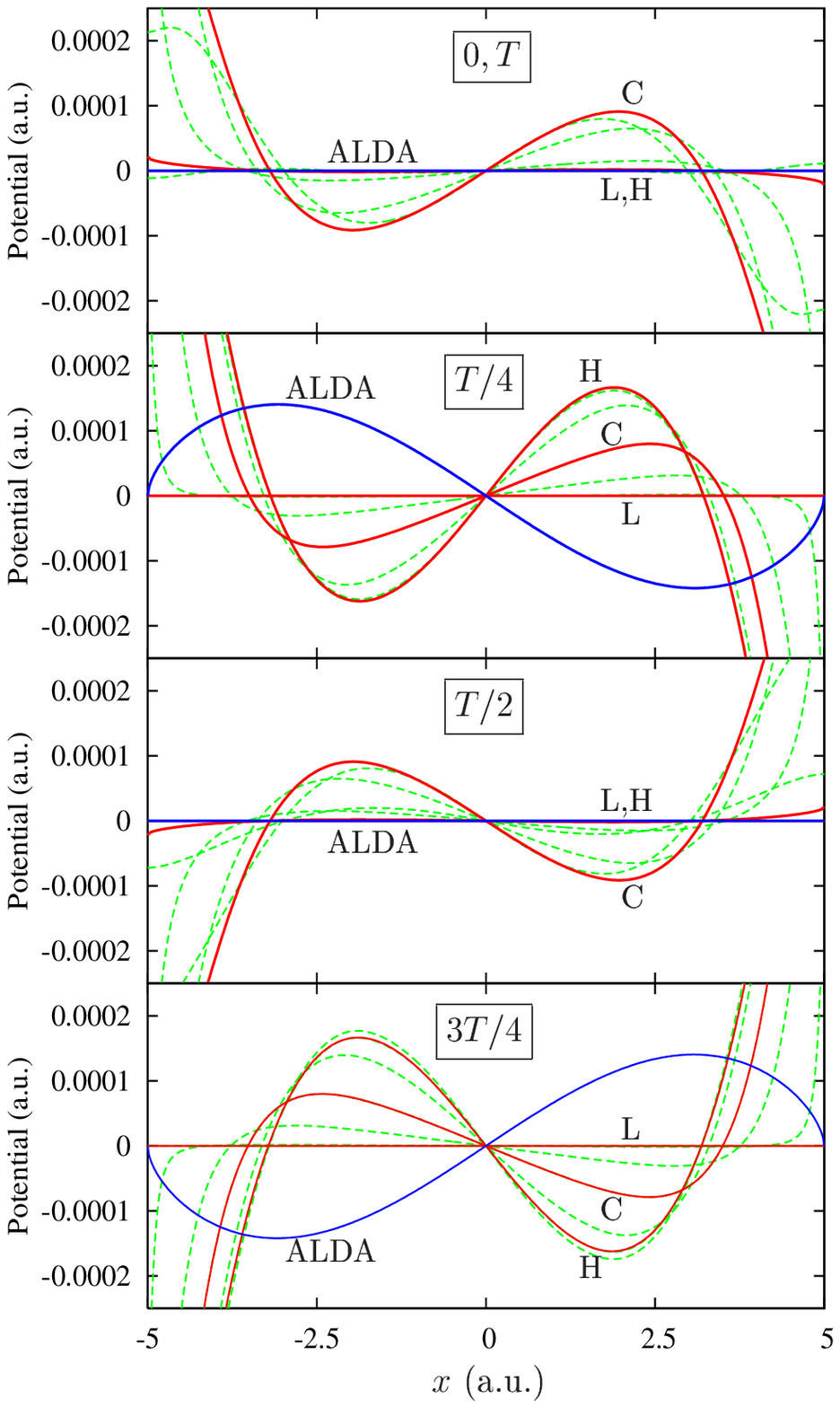}}}
\end{picture}
\caption{\label{figure4} (Color online) Snapshots of ALDA xc potential fluctuations and xc memory contributions
during one cycle of the sloshing mode
(Fig. \ref{figure2}), with amplitude $A=0.005$. Blue solid lines: $\widetilde{V}_{\rm xc}^{\rm ALDA}$,
scaled by 0.1 (independent of frequency).
Red solid lines: $V_{\rm xc}^{\rm M}$, in GK parametrization, for low frequency, $\omega_{\rm L} = 0.001 \: \bar{\omega}_p$,
crossover frequency, $\omega_{\rm C} = 1.7\: \bar{\omega}_p$, and high frequency, $\omega_{\rm H} = 1000\: \bar{\omega}_p$.
Green dashed lines: $V_{\rm xc}^{\rm M}$ at $0.1, 0.5, 2, 10$ times $\omega_{\rm C}$.}
\end{figure}

\begin{figure}
\unitlength1cm
\begin{picture}(5.0,13.25)
\put(-9.25,-10.25){\makebox(5.0,13.25){
\includegraphics{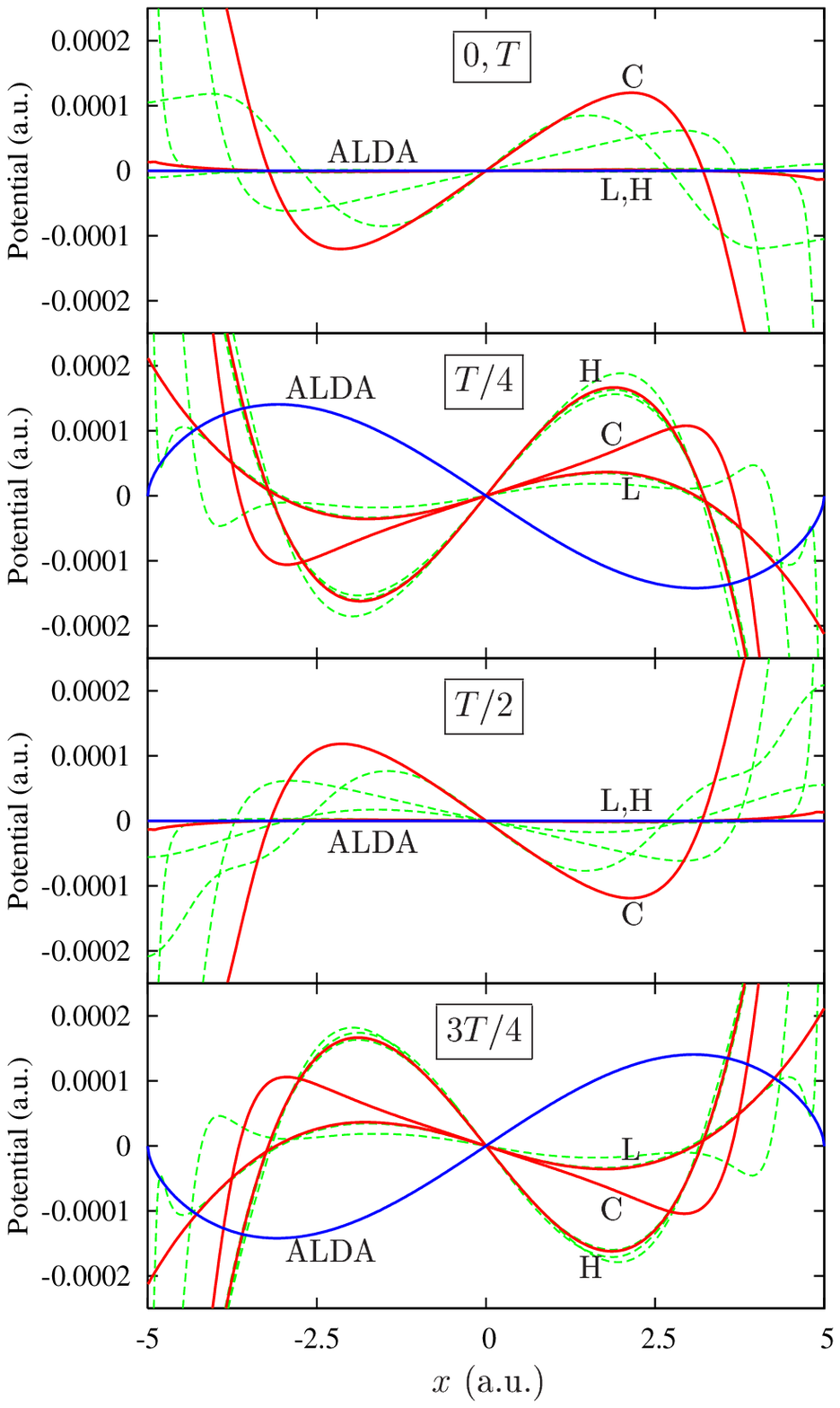}}}
\end{picture}
\caption{\label{figure5} (Color online) Same as Fig. \ref{figure4}, but
$V_{\rm xc}^{\rm M}$ calculated in QV parametrization. The crossover frequency is $\omega_{\rm C}=2.15\: \bar{\omega}_p$.}
\end{figure}

\begin{figure}
\unitlength1cm
\begin{picture}(5.0,13.25)
\put(-9.25,-10.25){\makebox(5.0,13.25){
\includegraphics{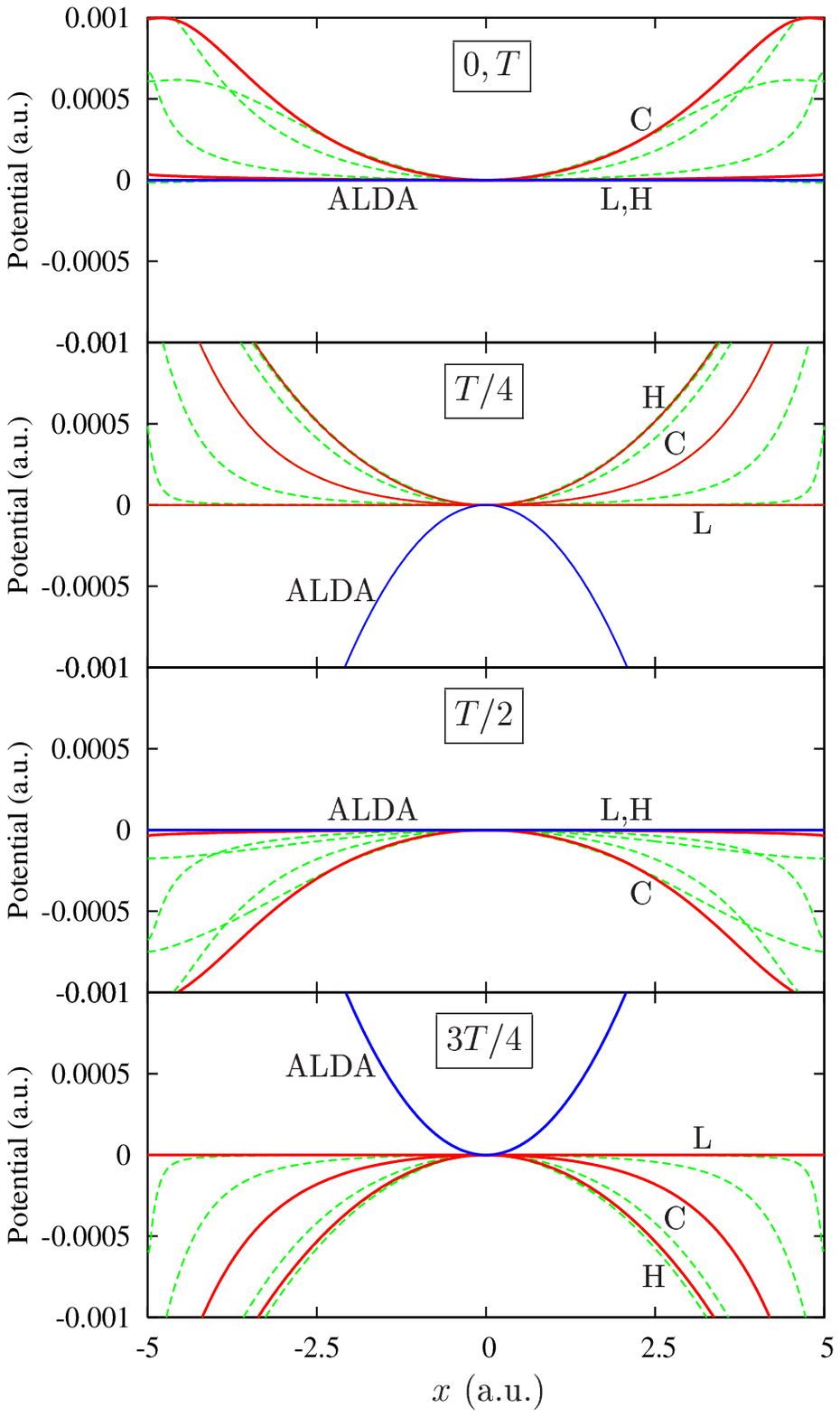}}}
\end{picture}
\caption{\label{figure6} (Color online) Same as Fig. \ref{figure4}, but
for the breathing mode (Fig. \ref{figure3}).  $\widetilde{V}_{\rm xc}^{\rm ALDA}$ is not scaled.
The crossover frequency is $\omega_{\rm C}=2.22\: \bar{\omega}_p$}
\end{figure}

\begin{figure}
\unitlength1cm
\begin{picture}(5.0,13.25)
\put(-9.25,-10.25){\makebox(5.0,13.25){
\includegraphics{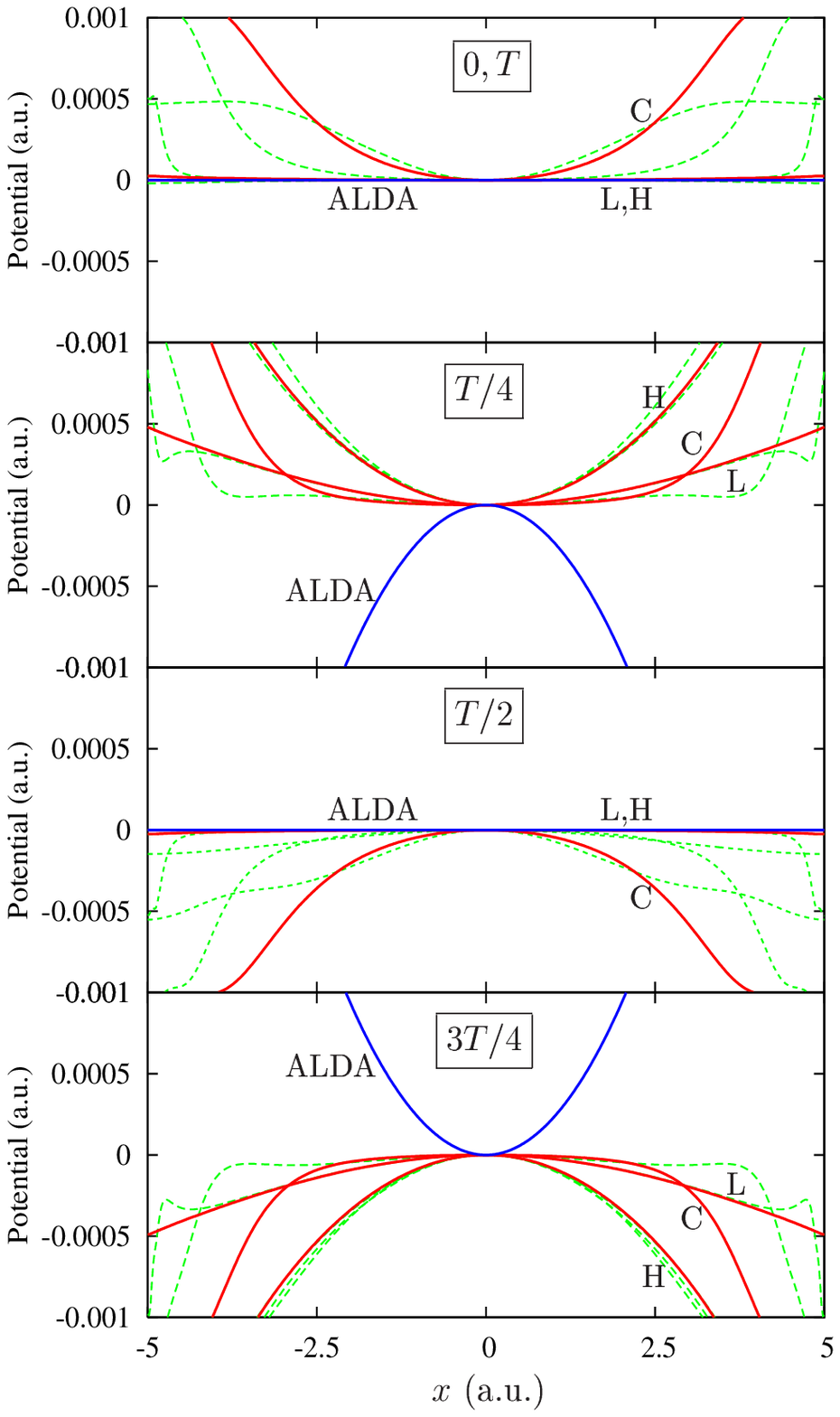}}}
\end{picture}
\caption{\label{figure7} (Color online) Same as Fig. \ref{figure6}, but
$V_{\rm xc}^{\rm M}$ calculated in QV parametrization. The crossover frequency is $\omega_{\rm C}=2.65\: \bar{\omega}_p$}
\end{figure}

Figures \ref{figure4}--\ref{figure7} each show four snapshots of $V_{\rm xc}^{\rm M}(x,t)$ and
$\widetilde{V}_{\rm xc}^{\rm ALDA}$ during one cycle of the sloshing/breathing modes, at
$t=0,T$ and $T/2$ (density passes through equilibrium, with maximal velocity) and
$t=T/4$ and $3T/4$ (density at turning points), as illustrated in Figs. \ref{figure2},\ref{figure3}.
$V_{\rm xc}^{\rm M}(x,t)$ has been calculated using the GK and QV parametrizations as input
(Figs. \ref{figure4},\ref{figure6} and Figs. \ref{figure5},\ref{figure7}, respectively).
All density oscillations have the same small amplitude $A=0.005$, but different frequencies.
We highlight the low- and high-frequency limit, $\omega_{\rm L} = 0.001 \: \bar{\omega}_p$ and
$\omega_{\rm H} = 1000 \: \bar{\omega}_p$,
respectively, and an intermediate crossover frequency $\omega_{\rm C}$, to be defined below, which varies between
$1.7 \: \bar{\omega}_p$ and $2.65 \: \bar{\omega}_p$ for the four cases considered. To illustrate
the strong dependence on frequency in the crossover regime, we also show $V_{\rm xc}^{\rm M}$ at
$0.1, 0.5, 2$, and 10 times $\omega_{\rm C}$. In the high-frequency limit $\omega_{\rm H}$,
$V_{\rm xc}^{\rm M}(x,t)$ is identical to $\widetilde{V}_{\rm xc}^{\rm E}(x,t)$ in all cases, i.e.,
the dynamics is purely elastic.

The results in Figs. \ref{figure4}--\ref{figure7} reveal the following features:

(i) The high-frequency limit of $V_{\rm xc}^{\rm M}(x,t)$
[i.e., $\widetilde{V}_{\rm xc}^{\rm E}(x,t)$] is phase-shifted
by $\pi$ with respect to $\widetilde{V}_{\rm xc}^{\rm ALDA}(x,t)$
in all cases considered. This is to be expected for a purely elastic potential: it reaches
its maximum at the instant of largest displacement from equilibrium.

(ii) At the low-frequency limit $\omega_{\rm L}$, the behavior of $V_{\rm xc}^{\rm M}$
depends on the parametrization used to calculate the memory kernel:
in GK, $V_{\rm xc}^{\rm M}$ vanishes, whereas in QV, $V_{\rm xc}^{\rm M}$ again becomes
purely elastic. This reflects the different long-time behaviors of the memory kernel:
in GK, it decreases exponentially, whereas in QV it approaches a finite constant (see Fig. \ref{figure1}).

(iii) At intermediate frequencies, the phase shift between $\widetilde{V}_{\rm xc}^{\rm ALDA}$
and $V_{\rm xc}^{\rm M}$ varies between $\pi$ and $\pi/2$, where $\pi/2$ indicates a purely dissipative
potential (see below).

(iv) The average strength of the forces associated with $V_{\rm xc}^{\rm M}$ grows with frequency,
and becomes comparable to the ALDA fluctuating forces in the high-frequency limit (notice that
$\widetilde{V}_{\rm xc}^{\rm ALDA}$ is scaled by 0.1 in Figs. \ref{figure4} and \ref{figure5}).
This clearly shows that non-adiabatic effects can become non-negligible in practice. We will
say more about this below when we discuss the nonlinear regime.

\subsubsection{Power and dissipation}

\begin{figure}
\unitlength1cm
\begin{picture}(5.0,8.5)
\put(-8.0,-11.8){\makebox(5.0,8.5){\includegraphics{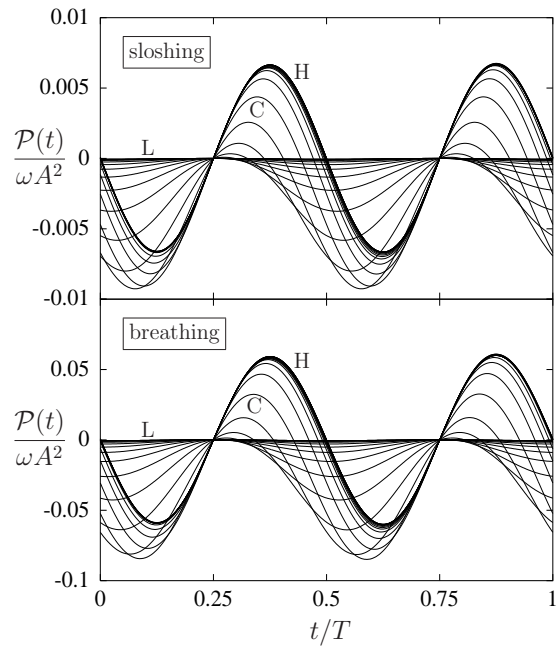}}}
\end{picture}
\caption{\label{figure8} Power ${\cal P}(t)$ [Eq. (\ref{power})] calculated using GK, associated with the sloshing
and the breathing mode, during one
cycle of the charge-density oscillation $(T=2\pi/\omega)$, for different frequencies. L,C,H indicate the
low-frequency (purely dissipative in GK), crossover, and high-frequency regime (purely elastic).
}
\end{figure}

For a more quantitative analysis, it is useful to consider the power associated with
the charge-density oscillations of the two types of modes. We define the power in the usual way as
force density times velocity:
\begin{equation} \label{power}
{\cal P}(t) =  \int dx \: v(x,t)\: n(x,t) \:\frac{\partial}{\partial x} V_{\rm xc}^{\rm M}(x,t) \:.
\end{equation}
Figure \ref{figure8} shows ${\cal P}(t)$, scaled by $\omega A^2$, during one cycle of the sloshing and breathing modes
(calculated using the GK parametrization). In the low-frequency limit, where the currents are vanishing,
the power tends to zero, but as the frequency increases, a periodic input/output of power is observed to
take place during a cycle. It can be clearly seen that, on average, $\cal P$ is more negative than positive
for intermediate frequencies, which indicates net power dissipation. In the high-frequency, elastic limit,
${\cal P}(t)$ has sizable amplitudes, but averages to zero during a cycle.

\begin{figure}
\unitlength1cm
\begin{picture}(5.0,8.5)
\put(-8.0,-11.8){\makebox(5.0,8.5){\includegraphics{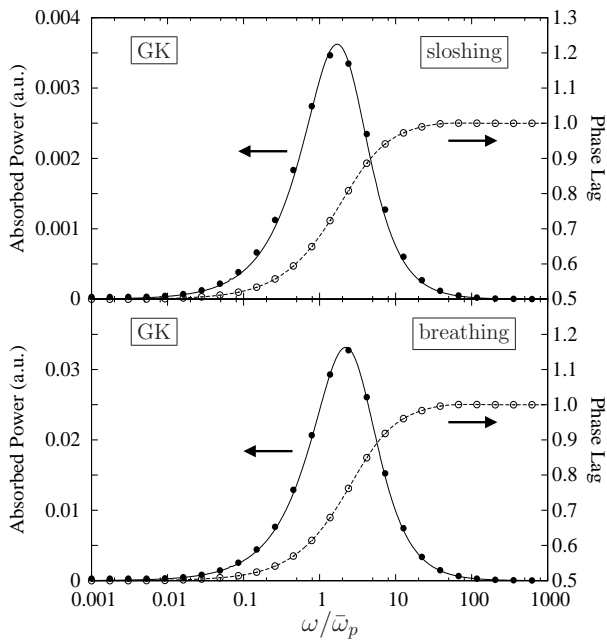}}}
\end{picture}
\caption{\label{figure9} Full symbols: net rate of power absorption, $-\bar{\cal P}/\omega A^2$, during one
cycle of the sloshing and breathing modes, versus frequency (in GK). Full line: $- \gamma \Im f_{\rm xc}^L(\omega)$,
for $r_s=1.41$ (sloshing) and 1.18 (breathing).
Open symbols: Phase lag (in units of $\pi$) between ${\cal P}(t)$ and ALDA.
The crossover frequency is defined by a maximum in power absorption and a phase lag $3\pi/4$.
The dashed lines are a guide to the eye.
}
\end{figure}

\begin{figure}
\unitlength1cm
\begin{picture}(5.0,8.5)
\put(-8.0,-11.8){\makebox(5.0,8.5){\includegraphics{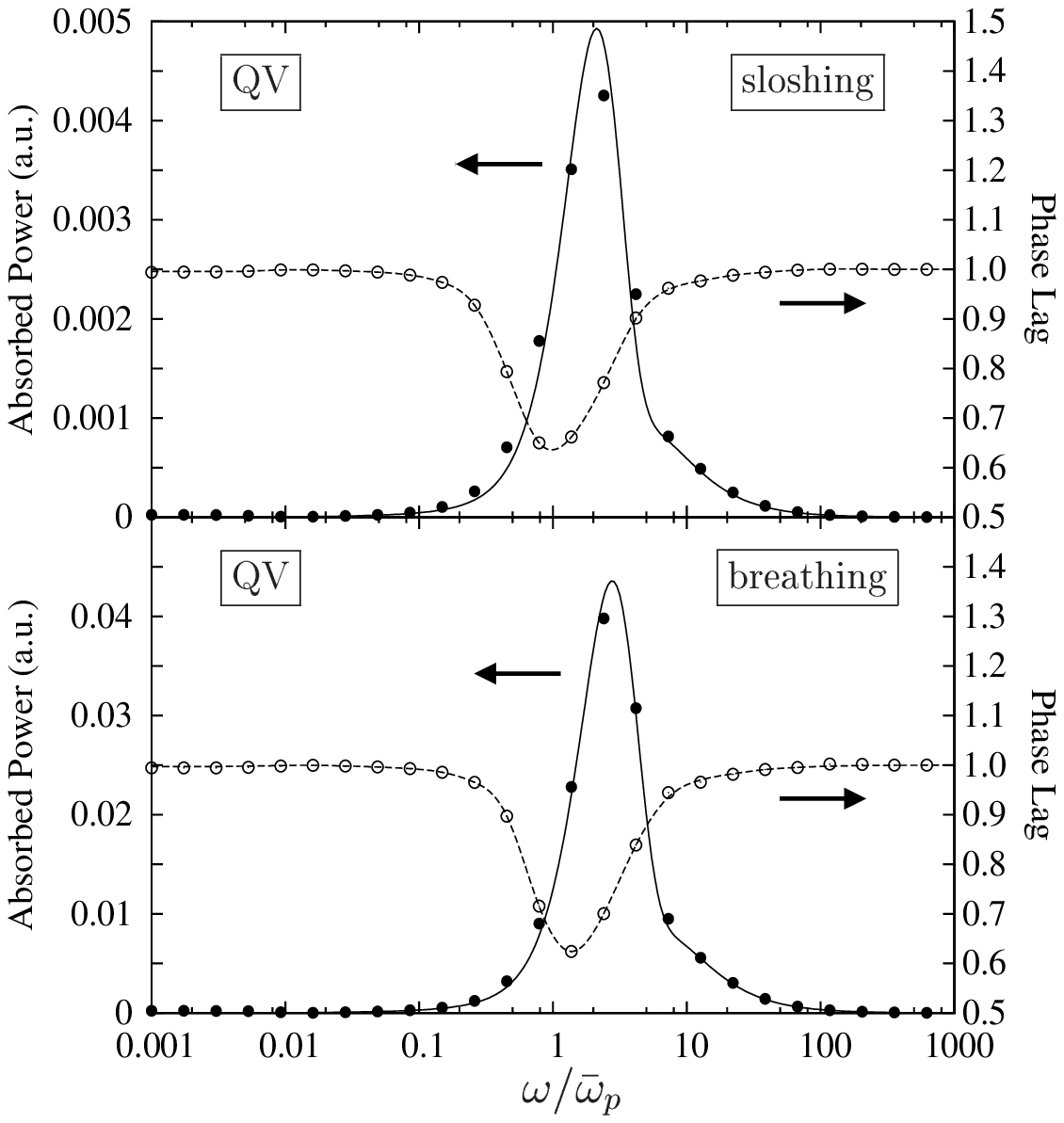}}}
\end{picture}
\caption{\label{figure10} Same as Fig. \ref{figure9}, but in QV parametrization. Both the low- and
high-frequency limits are purely elastic.}
\end{figure}

These findings are summarized in Figures \ref{figure9} and \ref{figure10}. We plot $-\bar{\cal P}$, the absolute value of
the time-average of ${\cal P}(t)$, scaled by $\omega A^2$, which represents the net power absorption per cycle, for
the sloshing and the breathing modes. Notice that $-\bar{\cal P}$ in all cases has a pronounced enhancement
for frequencies of the order of $\bar{\omega}_p$. We define the crossover frequencies $\omega_{\rm C}$ as
those frequencies where the maxima of power absorption occur.

While the two modes behave qualitatively very similarly as far as their power absorption is concerned,
one observes in Figs. \ref{figure8}--\ref{figure10}
that the absorbed power of the breathing mode is about an order of magnitude higher than for
the sloshing mode, for the same value of the amplitude, $A=0.005$. This hardly comes as a surprise:
our sloshing mode can be viewed as a cousin of Kohn's mode, \cite{kohn1961,BreyYip} and the Harmonic Potential
Theorem \cite{dobson1994} tells us that charge-density oscillations in parabolic quantum wells are undamped.
The breathing mode, on the other hand, bears no resemblance at all to Kohn's mode. The sloshing mode has a
smooth hydrodynamic flow with relatively little internal compression except at the turning points.
By contrast, the defining feature of the breathing mode is the periodic compression and rarefaction of its
density profile, i.e., a very ``un-hydrodynamic'' behavior.

It turns out that there is a direct relation between the average power absorption and
the imaginary part of the xc kernel, $\Im f_{\rm xc}^L(\omega)$. This is clear from Eq. (\ref{Ykernel}),
which expresses the memory kernel $Y(n,t-t')$ via Fourier transform of $\Im f_{\rm xc}^L(\omega)/\omega$.
The full lines in Figs. \ref{figure9} and \ref{figure10} plot $- \gamma \Im f_{\rm xc}^L(\omega)$, where
$\gamma$ is a constant scaling factor. For a best fit, $\gamma_{\rm sl}=0.00525$
and $r_s^{\rm sl}=1.41$ for the sloshing mode, and $\gamma_{\rm br}=0.071$ and $r_s^{\rm br}=1.18$ for the breathing mode
(same for GK and QV). Notice that the equilibrium density $n_0(x)$ has a value of $r_s=1.06$ in the center.
It thus emerges that the dissipation is dominated by different regions of the density distribution for
the two modes: around $x=\pm 0.27 L$ for the sloshing mode, i.e., halfway between center and walls,
and $x=\pm 0.18 L$ for the breathing mode, i.e., much closer to the center.

The phase lag of the modes as a function of frequency was already discussed
in the context of Figs. \ref{figure4}-\ref{figure7}. However, in Figs. \ref{figure9} and \ref{figure10}
one can see the difference between GK and QV  most dramatically.
We consider the phase lag between the instantaneous power absorption, ${\cal P}(t)$, and the ALDA
potential fluctuations (which are in phase with the charge-density oscillations).
Fig. \ref{figure9} shows that in GK one has a
transition from a purely dissipative behavior in the low-frequency limit, with phase difference $\pi/2$,
via a crossover region of mixed dissipative/elastic behavior, to the high-frequency, purely elastic
regime with phase lag $\pi$. The maximal power absorption occurs
for a phase lag of $3\pi/4$. On the other hand, in the QV parametrization the low-frequency regime is
also purely elastic, i.e. both the low- and high-frequency limits have phase lag $\pi$,
and dissipative contributions come in only at intermediate frequencies. The resulting power loss
as a function of frequency is thus a bit higher and more narrowly peaked about $\omega_{\rm C}$ in
QV than in GK.

As mentioned earlier, the qualitative differences between GK and QV have their origin
in the different long-time behavior of the memory kernels, see Fig. \ref{figure1} and the discussion
in Section \ref{sec:C-TDDFT-nonlinear}.
The overall result is that there is a broad range of frequencies, between about 0.1 to 10 times the
characteristic average plasma frequency $\bar{\omega}_p$, where the system exhibits a mixed elastic/dissipative
behavior, which can lead to substantial dissipation.

\subsection{Nonlinear regime} \label{nonlin}

\subsubsection{C- versus L-TDDFT in the high-frequency limit}

In the following, we will extend our numerical studies of C- and L-TDDFT into the nonlinear
regime. We begin by directing our attention to the high-frequency, purely elastic region, since our approximate
version of  L-TDDFT becomes exact in that limit and can thus be used as a benchmark to assess the accuracy of nonlinear C-TDDFT.

\begin{figure}
\unitlength1cm
\begin{picture}(5.0,9)
\put(-8.0,-12.25){\makebox(5.0,9){\includegraphics{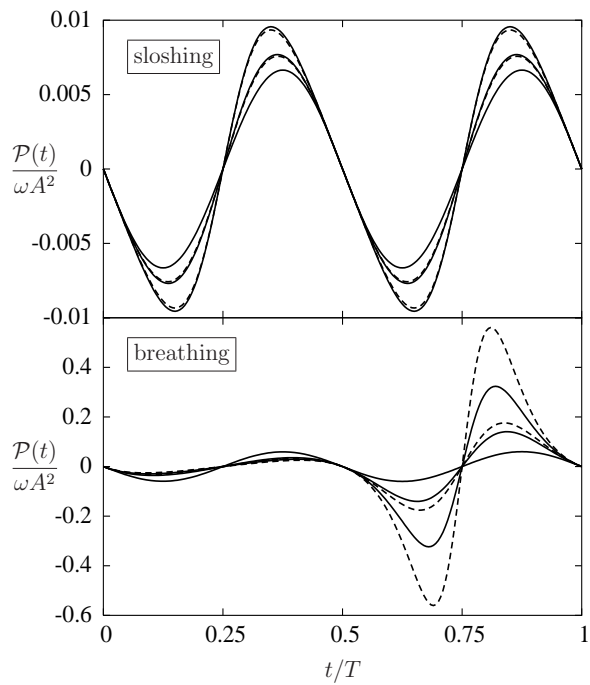}}}
\end{picture}
\caption{\label{figure11} Power ${\cal P}(t)$ [Eq. (\ref{power})] calculated using GK, over one cycle of the sloshing
and the breathing modes, for $\omega = 10^4 \tilde{\omega}_p$. Full and dashed lines: C- and L-TDDFT for amplitudes
$A=0.005$, $0.5$, and $0.75$. At $A=0.005$, C- and L-TDDFT are identical (see Fig. \ref{figure8}).
}
\end{figure}

Fig. \ref{figure11} shows the scaled power ${\cal P}(t)/\omega A^2$ [Eq. (\ref{power})] for
$\omega = 10^4 \tilde{\omega}_p$ during one cycle of the sloshing and breathing modes, comparing small amplitudes
($A=0.005$) and large amplitudes ($A=0.5$ and $0.75$).
In the small-amplitude regime, ${\cal P}(t)/\omega A^2$ has a sinusoidal shape for both modes,
independent of the amplitude $A$ of the oscillations (see also Fig. \ref{figure8}),
but for large amplitudes, nonlinear deviations occur.
In the sloshing mode, ${\cal P}(t)$ tends to a more sawtooth-like shape.
In the breathing mode, ${\cal P}(t)$ becomes suppressed when the charge density spreads out in the first
half of the cycle, and dramatically enhanced when the density gets squeezed in the second half of the cycle.
As we will see below, these strong deformations give rise to sizable non-adiabatic effects.

For small amplitude, $A=0.005$, C- and L-TDDFT are identical, but for larger amplitudes, $A=0.5$ and $0.75$,
some differences develop. However, the magnitude of these deviations strongly depends
on the type of mode. For the sloshing mode, we find that C-TDDFT remains quite close to L-TDDFT
even for large-amplitude oscillations, whereas in the breathing mode, C-TDDFT deviates from
L-TDDFT by about a factor of two for large values of $A$.

\begin{figure}
\unitlength1cm
\begin{picture}(5.0,8.25)
\put(-8.25,-12.5){\makebox(5.0,8.25){\includegraphics{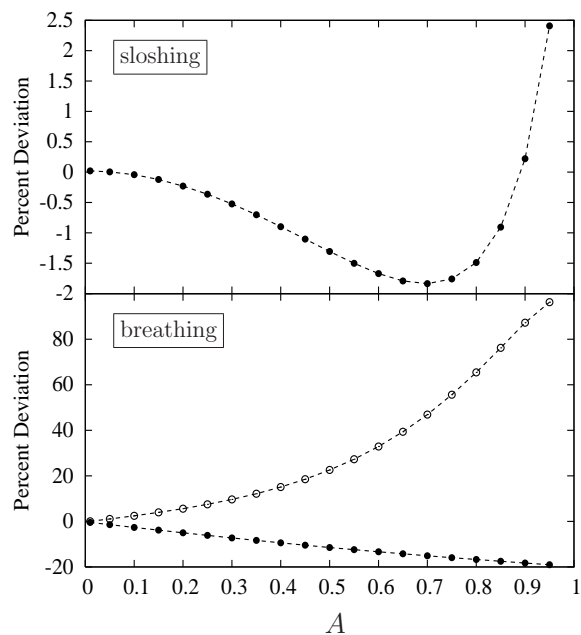}}}
\end{picture}
\caption{\label{figure12} Percent deviation of C-TDDFT from the (exact) L-TDDFT result for the time-averaged
absolute value of the scaled power,
$(t_2-t_2)^{-1}\int_{t_1}^{t_2} dt|{\cal P}(t)|/\omega A^2$. Top panel: sloshing mode, averaged over
one cycle. Bottom panel: breathing mode,
averaged over the first (full symbols) and the second half-cycle (open symbols).
}
\end{figure}

A more quantitative measure of the deviation of C-TDDFT from the exact high-frequency L-TDDFT result is shown in
Fig. \ref{figure12}. We plot the difference in percent of the time average of the absolute value of the scaled power,
$(t_2-t_2)^{-1}\int_{t_1}^{t_2}dt|{\cal P}(t)|/\omega A^2$, over one cycle for the sloshing mode, and over the first and second
half-cycle for the breathing mode, as a function of the amplitude $A$. As expected, the difference is seen to be increasing
as the amplitude grows, but the two modes exhibit a very different behavior.

Due to its close kinship to Kohn's mode, it comes as no surprise that the deviation is much
smaller for the sloshing mode, namely at most around 2 \% even for very large amplitudes.
The C-TDDFT error of the breathing mode is larger by at least an order of magnitude.
However, the overall error of C-TDDFT in the high-frequency limit compared to L-TDDFT remains surprisingly small,
as long as the amplitude is not too large: For $A=0.2$, we get a deviation of about 0.2 \% for the sloshing mode,
and about 5 \% for the breathing mode. For the largest amplitude considered ($A=0.9$),
we obtain a deviation of 2.5 \% for the sloshing mode, and 20 \% and 100 \% for first and second half-cycles of
the breathing mode. In the latter case, the deformations are so strong as to invalidate the basic assumptions
used to derive the simple form of nonlinear C-TDDFT.

\subsubsection{Non-adiabatic corrections to the ALDA}

The preceding high-frequency analysis shows that the C-TDDFT non-adiabatic xc potentials
remain close (to within a few percent) to the exact L-TDDFT results, except for modes with extremely strong deformations
such as sloshing modes with $A \agt 0.9$ or breathing modes with $A \agt 0.2$. As long as the deformations
remain within these approximate limits, it is reasonable to expect C-TDDFT to be accurate for finite frequencies as well.

\begin{figure}
\unitlength1cm
\begin{picture}(5.0,13.25)
\put(-9.25,-10.25){\makebox(5.0,13.25){
\includegraphics{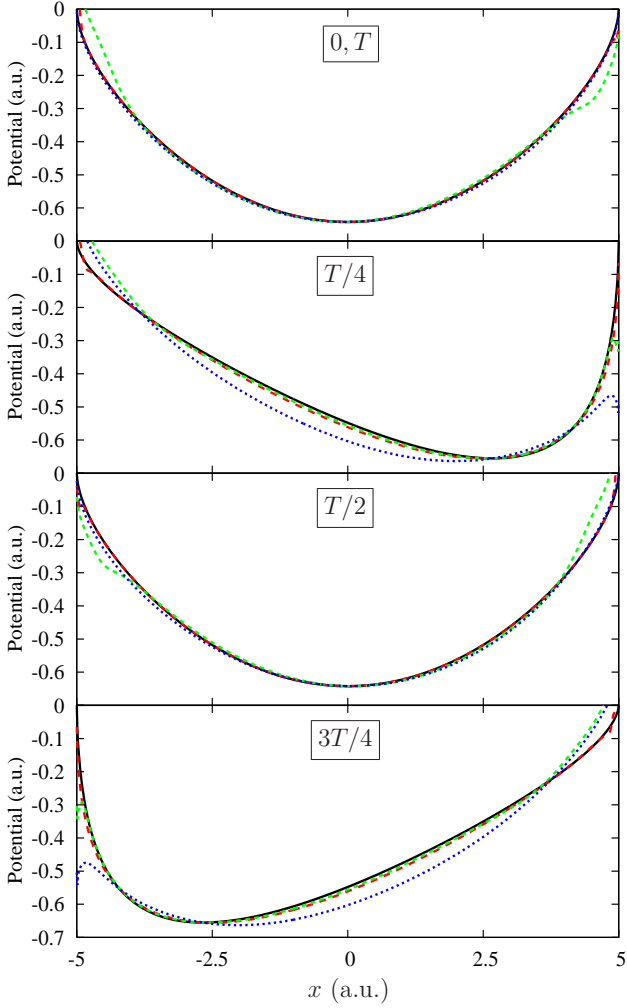}}}
\end{picture}
\caption{\label{figure13} (Color online) Snapshots of the adiabatic and non-adiabatic xc
potentials during one cycle of the sloshing mode, with amplitude $A=0.75$. Black solid line:
$V_{\rm xc}\upalda$. Red long-dashed, green medium-dashed, and blue dotted lines:
$V_{\rm xc}\upalda+V_{\rm xc}^{\rm M}$ for frequencies $\omega =0.1 \tilde{\omega}_p$,
$\omega =\tilde{\omega}_p$, and $\omega =10\tilde{\omega}_p$.}
\end{figure}

\begin{figure}
\unitlength1cm
\begin{picture}(5.0,13.25)
\put(-9.25,-10.25){\makebox(5.0,13.25){
\includegraphics{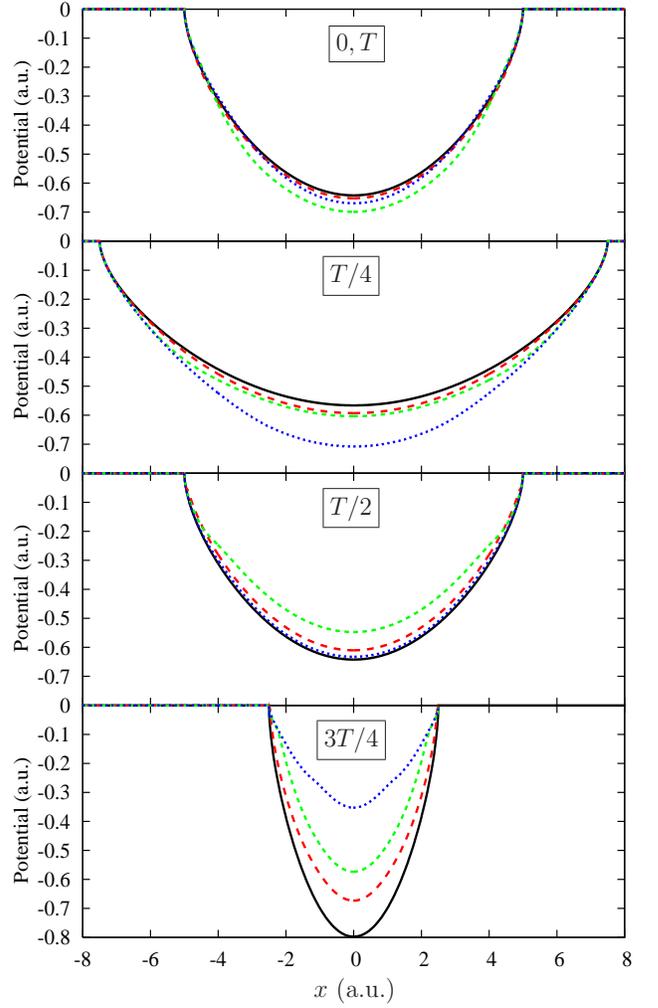}}}
\end{picture}
\caption{\label{figure14} (Color online) Same as Fig. \ref{figure13}, but
for the breathing mode with amplitude $A=0.5$.}
\end{figure}

To get an impression of the magnitude of the non-adiabatic corrections to the ALDA,
we plot in figures \ref{figure13} and \ref{figure14} the full adiabatic and non-adiabatic xc potentials,
$V_{\rm xc}\upalda(x,t)$ and $V_{\rm xc}\upalda(x,t)+V_{\rm xc}^{\rm M}(x,t)$, where $V_{\rm xc}^{\rm M}$
is calculated within C-TDDFT using the QV parametrization. Fig. \ref{figure13} shows results for
a large-amplitude sloshing mode with $A=0.75$, and Fig. \ref{figure14} for a breathing mode with $A=0.5$.
The figures show four snapshots taken during one cycle of the mode, similar to what was presented
in Figs. \ref{figure4}--\ref{figure7}, except that now we are plotting the total potential.
The ALDA xc potential is of course independent of the mode frequency,
and we compare it with the ALDA+M potential at three different frequencies: $\omega =0.1 \tilde{\omega}_p$ (low-frequency regime),
$\omega =\tilde{\omega}_p$ (around cross-over), and $\omega =10\tilde{\omega}_p$ (approaching the high-frequency regime).

The results in Figs. \ref{figure13} confirm again that the non-adiabatic effects for the sloshing
mode are relatively modest, even for large-amplitude deformations. The post-ALDA corrections
become more important for the high-frequency oscillations ($\omega =10\tilde{\omega}_p$), where one finds deviations from ALDA
of the order of 10\% at the turning points of the oscillation at $T/4$ and $3T/4$. For the lower-frequency
modes, the non-adiabatic corrections to the ALDA stay mostly within about 1\%.

On the other hand, the breathing mode exhibits much more dramatic non-adiabatic effects, see Fig. \ref{figure14}.
Again we find that the post-ALDA corrections are moderate for the lower frequencies considered. However,
for the high-frequency case ($\omega =10\tilde{\omega}_p$), we find that at the instances of maximum
deformation ($T/4$ and $3T/4$) the memory effects cause a correction to the ALDA of up to a factor of 2, which is indeed
substantial.

The impact of the high-frequency post-ALDA corrections is similar for both modes: they tend to oppose the
deformation of the ALDA potential at the instances of maximum deformation of the density distribution.
For the breathing mode, this means that $V_{\rm xc}$ becomes less broad at $T/4$ and less deep at $3T/4$,
and for the sloshing mode, that the potential minimum lies closer to the center, and is less sharp.
In a more realistic calculation where the density, instead of being a given function, follows from solving a
TDKS calculation with a time-dependent driving potential, this would imply that the elasticity of the electron
liquid tends to counteract deformations of the density, making the system a more rigid and
somewhat harder to deform.

\begin{figure}
\unitlength1cm
\begin{picture}(5.0,9.9)
\put(-8.75,-10.6){\makebox(5.0,9.9){
\includegraphics{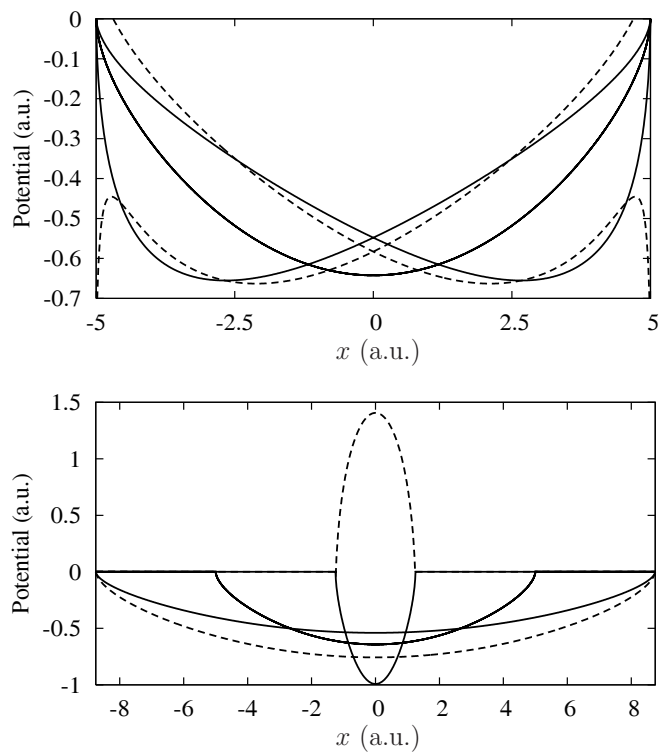}}}
\end{picture}
\caption{\label{figure15} Comparison of ALDA (full lines) and L-TDDFT (dashed lines) xc potentials for the
sloshing mode (top panel) and the breathing mode (bottom panel), both with amplitude $A=0.75$.
The same snapshots during one cycle are shown as in figures \ref{figure13} and \ref{figure14}. }
\end{figure}

Lastly, to illustrate a case of extreme non-adiabaticity
let us consider the high-frequency limit of strongly nonlinear dynamics. Fig. \ref{figure15} shows a comparison of the
ALDA and L-TDDFT xc potentials for both modes with amplitude $A=0.75$, for the same snapshots taken during one cycle
as previously in Figs. \ref{figure13} and  \ref{figure14}. For the instants $0$, $T/2$ and $T$,
ALDA and L-TDDFT coincide, but for $T/4$ and $3T/4$, large deviations occur. The most dramatic
non-adiabatic effect is observed for the breathing mode at $3T/4$, where the elastic contribution
of L-TDDFT is so large that the resulting xc potential is of the same magnitude as the ALDA, but
with opposite sign. This clearly shows that in situations in which the electron density is
rapidly and strongly deformed the ALDA becomes drastically wrong, with non-adiabatic
corrections of the same order of magnitude as the potential itself.

\section{Conclusion} \label{sec:conclusion}

In this paper, we have carried out a detailed comparative study of two non-adiabatic local approximations
for the xc potential in TDDFT, using quasi-one-dimensional model systems (whose electronic density is uniform
in two spatial directions, and nonuniform in the third). The goal was to compare the
xc potentials resulting from C-TDDFT and L-TDDFT when a given time-dependent density is
used as input, and to determine the magnitude of the resulting non-adiabatic effects
when compared to the ALDA. These comparisons were carried
out for two types of collective modes: a sloshing and a breathing mode, for a wide dynamical
range of amplitudes and frequencies.

In our discussion of non-adiabatic xc effects, we were paying particular attention to
dynamical regimes where the two theories, C-TDDFT and elastic L-TDDFT, are known to be exact:
C-TDDFT in the linear regime of small-amplitude oscillations, regardless of the frequency,
and elastic L-TDDFT in the high-frequency limit, for oscillations of an arbitrary amplitude.
This helps to shed light on the so far open question concerning the validity of
C-TDDFT in the nonlinear regime: it turns out that, at least in the high-frequency
limit, C-TDDFT is surprisingly accurate. Our results show that for
moderate deformations (up to 20\% of the initial density distribution) errors in the xc potential remain within a few percent.
This gives us good reason to believe that this nonlinear formalism should work well for finite frequencies, too (as long as
the deformations are not too large).

Let us now address the fundamental question of the meaning and the importance of ``non-adiabatic effects''
in TDDFT, based on the insights resulting from our study. In analyzing these effects, the language of
hydrodynamics, as it is used in C- and L-TDDFT, proves very useful. In general, non-adiabaticity
manifests itself through elastic and dissipative components of the electron dynamics, which can be distinguished by their
characteristic phase lag with respect to the ALDA. Both elastic and
dissipative effects enter the theory via the dependence of xc
potential on the deformation tensor. Since the deformation is defined
relative to the initial state, this dependence reflects how much of its history the system carries in its memory.
A local-in-time dependence on the deformation, which occurs in the high-frequency regime, corresponds to
an extremely pronounced memory and a purely elastic xc potential. The dissipative contribution formally appears in the form of
a time-nonlocality in the dependence of xc potential on deformations of the electron fluid. In general this time-nonlocality
tends to shorten the characteristic memory time, and, as a result, it somewhat diminishes the elastic contribution. In addition
it brings about fundamentally new effects, such as relaxation and the corresponding power absorption.  Which of the the two
contributions (elastic or dissipative) is dominant, or
whether both play an equally important role, depends on the dynamical regime in which the system under study
is evolving. The results from our simple model system lead us to the following conclusions:

{\em Linear regime.} For small-amplitude oscillations, dissipation is the most important and dramatic
consequence of non-adiabaticity, which leads to qualitative corrections to the adiabatic dynamics. There is a dynamical range
which we call ``crossover'' regime
in which the power absorption due to xc retardation effects is maximal. This crossover regime
occurs for frequencies that are comparable to an average plasmon frequency for the system.
On the other hand, for very low or for very high frequencies, dissipation vanishes. Elastic effects
in the linear regime appear to be less important, in a sense that they do not qualitatively change an overall behavior of the
xc potential. However, the quantitative effect of elastic corrections to the linearized ALDA can be significant, especially
in the high-frequency regime. In particular, for our breathing mode the post-ALDA non-adiabatic corrections are in general
of the same order of magnitude as the dynamic part of the ALDA potential itself.

{\em Nonlinear regime.} Here, elastic effects become more important, especially at high frequencies,
which can lead to substantial, and in certain regimes absolutely dominant contributions to the time-dependent xc potential.
As we discussed for the sloshing and breathing modes, the elasticity of the electron liquid tends to oppose
attempts to subject the electron system to strong and rapid deformations. This general tendency also naturally explains
why C-TDDFT has been successful for molecular polarizabilities, which are greatly
overestimated in ALDA. \cite{faassen1} Dynamic polarization of
the system corresponds to a redistribution of the charge density,
i.~e., to the deformation of an electron subsystem, which causes a
counteracting xc force. This force is an intrinsically non-adiabatic
effect that is completely missing in ALDA.

Our results once again illustrate the special role of Kohn's mode in TDDFT. If the electron dynamics
sufficiently resembles Kohn's mode, as is the case for the sloshing mode considered here,
non-adiabatic effects are generally small. On the other hand, for electron dynamics involving
high degrees of compression, such as our breathing mode, the non-adiabatic ``corrections'' can
become several times larger than the ALDA potential itself, so that the ALDA completely
breaks down, leading to a qualitatively wrong behavior. A striking illustration of such a situation was shown in Fig.~\ref{figure15}.

Finally, most of the discussions in this paper, especially those related to dissipative effects, are relevant only for
non-adiabaticity in extended
systems, where purely electronic dissipation has a well-defined meaning. \cite{DAgosta}
The situation is less clear when one attempts to describe non-adiabatic effects in finite
systems such as atoms and molecules. \cite{ullrichburke} Here, the elasticity of the electron
liquid leads to small shifts of excitation energies, but one also obtains finite linewidths due to
dissipation, which is clearly an unwanted effect. The question how non-adiabatic xc potentials for
small systems should be constructed thus remains an open issue.

\acknowledgments
C. A. U. acknowledges support from NSF Grant No. DMR-0553485 and from Research Corporation.
We thank Giovanni Vignale for fruitful discussions.


\begin{thebibliography}{100}
\bibitem{footnote1}
The sign difference reflects an ambiguity in the definition of the
stress tensor in the continuum mechanics literature. The standard
formulation of
C-TDDFT \cite{VUC} follows the sign convention used in the
classical elasticity theory, where the divergence of the stress tensor
$\sigma_{ij}$ defines a force exerted on a small volume element by
surroundings parts of the body. In contrast, L-TDDFT in its original
formulation \cite{Tokatly1,Tokatly2} accepts another convention that
is more common in fluid mechanics and in microscopic theory. Here
the sign of the stress tensor $P_{ij}$ coincides with the sign of
momentum flow tensor. With this definition the trace of $P_{ij}$ is
equal to the local pressure, while its divergence gives the force exerted
by an infinitesimal volume element on a surrounding fluid.
\end{thebibliography}
\end{document}